\documentclass[journal]{IEEEtran}
\usepackage{amsmath,amsfonts}
\usepackage{algorithmic}
\usepackage{algorithm}
\usepackage{array}
\usepackage[caption=false]{subfig}
\usepackage{textcomp}
\usepackage{stfloats}
\usepackage{url}
\usepackage{verbatim}
\usepackage{graphicx}
\usepackage{cite}
\usepackage{breqn}
\usepackage{amssymb}
\usepackage{placeins}
\begin{document}

\title{Unified Analytical Framework for SPAD Array Receivers with Dead-Time-Induced Blocking Loss and Inter-Symbol Interference in PAM-OWC Systems}

\author{Chen Wang, Zhiyong Xu, Jingyuan Wang, Jianhua Li, Weifeng Mou, Huatao Zhu
\thanks{This work was supported by National Natural Science Foundation of China (Grant No. 62271502, 62171463, 62471472) Hubei Provincial Natural Science Foundation of China (Grant No. 2026AFA101), and Natural Science Foundation of Jiangsu Province (Grant No. BK20231486). (\textit{Corresponding author: Jingyuan Wang}) 

Chen Wang, Weifeng Mou, Huatao Zhu are with the Information Support Force Engineering University, Wuhan 430010, China (e-mail: 0910210239@njust.edu.cn; weifengmou@126.com; zhuhuatao2008@163.com).

Zhiyong Xu, Jingyuan Wang,  and Jianhua Li are with the College of Communications Engineering, Army Engineering University of PLA, Nanjing 210007, China (e-mail: njxzy123@163.com; 13813975111@163.com; 18021528752@163.com).
}
\thanks{Manuscript received ; revised }}

\markboth{Journal of \LaTeX\ Class Files,~Vol.~xx, No.~xx, May~2026}%
{Shell \MakeLowercase{\textit{et al.}}: A Sample Article Using IEEEtran.cls for IEEE Journals}

\IEEEpubid{}

\maketitle

\begin{abstract}
Optical wireless communication (OWC) leveraging single-photon avalanche diode (SPAD) arrays offers exceptional sensitivity for photon-starving links. However, the inherent dead time of SPADs critically limits achievable data rates by introducing non-linear photon-counting distortions: blocking loss within a symbol duration and inter-symbol interference (ISI) across durations. This paper proposes a unified analytical framework capturing both distortions across all operational speed regimes for pulse-amplitude modulation (PAM), by establishing comprehensive statistical models for SPAD array receivers. For low and medium-speed systems (symbol duration longer than dead time), we derive exact closed-form expressions for the photon counts probability distribution using renewal theory, explicitly incorporating blocking loss and ISI. For high-speed systems (symbol duration shorter than dead time), we develop a Markov chain model characterizing the steady-state operational states and integrate it with trigger probability to obtain the exact binomial photon counts distribution. Furthermore, we propose low-complexity, near-optimal threshold detection schemes based on these models. This work provides essential theoretical tools for designing and optimizing high-performance SPAD-based OWC systems employing PAM.
\end{abstract}


\section{Introduction}
\IEEEPARstart{I}{n} recent years, optical wireless communication (OWC) has attracted growing interest as a promising complementary technology to radio frequency (RF) systems \cite{ref1,ref2}. Compared to RF, OWC offers advantages including larger bandwidth, lower emitting power, and smaller antenna aperture. However, in long-distance photon-starving applications, optical signals may fall below the sensitivity threshold of conventional receivers and become obscured by thermal noise \cite{ref5}. Single-photon avalanche diodes (SPADs) are a more suitable choice for such scenarios. SPADs provide extremely high internal gain, effectively overcoming thermal noise and enabling single-photon detection \cite{ref6}. Owing to single-photon sensitivity and substantial gain, SPADs have facilitated rapid advances in numerous applications. These detectors can approach quantum-limited sensitivity for weak optical signals and have consequently garnered significant attention in OWC research \cite{ref7, ref8, ref9,ref10}.

SPADs are semiconductor p-n junction devices operating under a specific principle: when reverse-biased slightly above the breakdown voltage, a high electric field is established. A single electron-hole pair generated within this region can then trigger a self-sustaining avalanche, resulting in substantial internal gain and a measurable current pulse. This current rapidly increases until the avalanche is quenched by reducing the bias voltage to or below the breakdown threshold \cite{ref11}. Detection of a subsequent photon requires restoring the bias above breakdown. This reduction and restoration of bias voltage is managed by a quenching circuit. The quenching process imposes a finite recovery period, known as the "dead time," during which the SPAD is unresponsive to incident photons \cite{ref12}. This paper employs active quenching (AQ) to achieve high-speed optical links, resulting in a non-paralyzable dead time \cite{ref13}. Within this context, operating modes are classified based on symbol duration relative to the dead time: {\bf Low-speed system:} Symbol durations significantly longer than the dead time; {\bf Medium-speed system:} Symbol durations approximately several times longer than the dead time; {\bf High-speed system:} Symbol durations equal to or shorter than the dead time.

Dead time constitutes a significant limitation for achievable data rates in SPAD-based OWC systems \cite{ref14,ref15}. Driven by the demand for higher data rates, several recent studies have focused on mitigating dead time effects through the use of SPAD arrays \cite{ref16,ref17,ref18}. SPAD arrays offer greater resilience to counting losses induced by dead time and enable higher photon counts rates. From a communication theory perspective, understanding the impact of dead time on SPAD-based OWC system performance is crucial. Therefore, precise modeling of the statistical photon-counting behavior of photon-counting receivers, accounting for dead time effects, is essential.

\subsection{Related Works}
Photon-counting receivers typically characterize received signals through discrete photon counts. In such Poisson channels, the traditional additive white Gaussian noise (AWGN) model is inapplicable. Early studies often assumed an ideal photon-counting receiver, neglecting dead time effects, and focused on the discrete-time Poisson (DTP) channel model \cite{ref19,ref20}. However, realizing such an ideal receiver is challenging. In practice, dead time introduces non-linear distortion to the photon counts \cite{ref21}. This dead time results in photon counts loss, inducing additional detection error events and thereby imposing extra performance constraints on the communication system. Crucially, this loss mechanism differs fundamentally from photon loss due to signal power attenuation. Specifically, the dead time triggered by one successfully detected photon can obstruct the detection of subsequent photons \cite{ref22}. Consequently, practical photon-counting channels cannot be simply modeled as DTP channels with reduced signal power \cite{ref23}.

Recent studies \cite{ref24,ref25} have proposed practical photon-counting receivers for photon-limited OWC systems, accounting for dead-time-induced blocking loss. Nonetheless, the assumption of Poisson statistics for the counts remains inaccurate. Ref. \cite{ref26} introduced the paralyzable and non-paralyzable count rate models, the most established frameworks for estimating the count rate of single-photon detectors. While count rate is a valuable metric for assessing detector speed, it provides an incomplete picture of detected and lost photons, which is essential for evaluating communication system error performance \cite{ref27}. Ref. \cite{ref28} studied the statistical behavior of AQ and passive quenching (PQ) SPAD receivers, deriving the exact photon counts distribution under the short dead time approximation. Ref. \cite{ref29} extended this approach, characterizing an AQ-SPAD array for OWC applications by modeling the receiver as a discrete memoryless channel (DMC). 

Beyond blocking loss within a single time interval, dead time can also cause counting loss in neighboring symbol durations. Specifically, the dead time from the last detected photon in one duration may extend into subsequent durations, temporarily blocking the SPAD at the start of those intervals. This causes extra counting losses and distorts the outputted photon counts, a phenomenon we term inter-symbol interference (ISI) \cite{ref30}. Ref. \cite{ref31} studied ISI in practical photon-counting receivers due to dead time, proposing a novel demodulation technique reported to achieve sensitivity gains of approximately 7 dB over traditional photon-counting schemes. However, this model lacks the capability for optical link error performance assessment. Ref. \cite{ref29} investigated the dead time effects of high-speed on the photon counts statistics of AQ-SPAD arrays, modeling the SPAD as a DMC. However, it did not derive a closed-form expression for the steady-state distribution, which is critical for capturing ISI effects, resorting instead to Monte-Carlo simulations – an approach that complicates system parameter optimization. Ref. \cite{ref32} employed a Markov chain model to characterize a high-speed pulse-position modulation (PPM) system using superconducting nanowire single-photon detectors (SNSPDs), deriving analytical expressions for symbol transition probabilities to characterize symbol error rate (SER). However, this model is limited to constant-power modulations like PPM and is unsuitable for non-constant-power schemes such as pulse-amplitude modulation (PAM).

In summary, photon counts statistics under dead-time-induced blocking loss have been studied primarily using DMC models. A key assumption in these studies is the independence of photon detection behavior across time intervals. While reasonable for low-speed systems where dead time is significantly shorter than the optical signal duration and blocking loss dominates. These analytical models are typically constrained to scenarios with ultra-short dead time (less than one-tenth of the symbol duration). For high-speed systems, Ref. \cite{ref32} established a statistical model incorporating dead-time-induced ISI effects for PPM, where ISI dominates. However, this model is not applicable to PAM. In addition, research on medium-speed system, where both blocking loss and ISI effects are comparably significant, remains lacking.

For these reasons, we believe establishing a comprehensive analytical model for SPAD array-based OWC systems employing PAM is crucial. This model should be applicable across both ultra-short and long dead time regimes. By accurately characterizing the distinct effects of blocking loss and ISI effects, we aim to provide versatile guidelines for optimizing system design.

\subsection{Our Contributions}
In this study, we establish a mathematical framework to precisely model the photon-counting behavior of SPAD array-based receivers operating over consecutive time intervals. For low-speed and medium-speed systems, we apply renewal theory to derive exact closed-form expressions for the photon counts probability distribution, accounting for both dead time effects: blocking loss and ISI effects. In contrast, for high-speed systems, we develop a Markov chain model characterizing the steady-state distribution of the SPAD’s operational states. Furthermore, by integrating a SPAD trigger probability model, we establish an exact binomial statistical model for photon counts in high-speed operation.

Our analysis demonstrates that the photon-counting process under dead time constraints cannot be accurately approximated by a Poisson distribution. Moreover, the analytical models required for low-speed and high-speed regimes are fundamentally distinct. To the best of our knowledge, no prior analytical work has derived the exact statistical distribution of photon counts while simultaneously addressing both blocking loss and ISI effects. Although focused on SPAD detectors, this approach is applicable to various single-photon detectors with similar dead-time behavior, including SNSPDs and photomultiplier tubes (PMTs).

The key contributions are summarized as follows:

{\bf(1) Renewal Process Model for low-Speed and Medium-Speed Systems:} We develop a renewal process model for low- and medium-speed PAM systems. Incorporating both blocking loss and ISI effects, we derive the closed-form expression for the photon counts distribution. Moreover, using multinomial framework, we compute the exact probability mass function (PMF) of photon counts for SPAD array receivers.

{\bf(2) Markov Chain Model for High-Speed Systems:} We formulate a Markov chain model for high-speed systems where the dead time is an integer multiple of the PAM symbol period. Based on the counting process, we classify symbol detection behavior into three states (active, detected, quenched) and define the corresponding state transition graph. Integrating this with the SPAD trigger probability model, we derive the exact PMF of photon counts.

{\bf(3) Enhanced Threshold Detection:} To decrease the computational complexity of signal demodulation, we propose two closed-form threshold formulas tailored for low-speed and high-speed systems, respectively. Simulation results demonstrate that these thresholds achieve nearly optimal symbol error performance relative to maximum likelihood (ML) detection.

\subsection{Organization}
The remainder of this paper is organized as follows. Section \uppercase\expandafter{\romannumeral2} derives the probability distribution of photon counts for low-speed, medium-speed and high-speed systems. Section \uppercase\expandafter{\romannumeral3} provides the SER of SPAD array-based OWC systems using both ML detection and threshold detection. Section \uppercase\expandafter{\romannumeral4} presents numerical results and discussions. Concluding remarks are presented in Section \uppercase\expandafter{\romannumeral5}. Notations used are collated in Table \ref{tab_1}. 

\renewcommand{\arraystretch}{1.2}
\begin{table}[!t]
\centering
\caption{Notations and preliminaries} 
\label{tab_1}
\begin{tabular}{|>{\centering\arraybackslash}m{0.20\linewidth}|>{\centering\arraybackslash}m{0.70\linewidth}|} 
\hline
\textbf{Notation} & \textbf{Definition} \\
\end{tabular}
\begin{tabular}{|>{\centering\arraybackslash}m{0.20\linewidth}|>{\raggedright\arraybackslash}m{0.70\linewidth}|} 
\hline
$p_{\mathrm{d}}, N_{\mathrm{A}}$ & Photon detection efficiency, array scale \\ \hline
$T_{\mathrm{s}}, \tau_{\mathrm{d}}, \xi, k_{\mathrm{max}}$ & Symbol duration, dead time, dead time ratio, maximum photon counts \\ \hline
$\lambda_{\mathrm{s}}, \lambda_{\mathrm{b}}, \lambda_{\mathrm{d}}$ & Signal photon rate, background photon rate, dark carrier rate \\ \hline
$N(t)$ & Stochastic variable denoting the number of events in the time interval $(0,t)$ \\ \hline
$M, x_m$ & PAM Modulation order, Symbol information \\ \hline
$p(k,t), P(k,s)$ & Probability of detecting $k$ photons during time interval $(0,t)$, Laplace transform of $p(k,t)$ with respect to the variable $t$ \\ \hline
$f_k(t), F_k(s)$ & Probability of detecting $k$-th photon at time instant $t$ in the symbol duration, Laplace transform of $f_k(t)$ with respect to the variable $t$ \\ \hline
$\mathcal{L}(\cdot), \mathcal{L}^{-1}(\cdot)$ & Laplace transform, Inverse Laplace transform \\ \hline
$\Lambda(k)$ & Likelihood ratio for photon counts $k$ \\ \hline
$k_{\mathrm{th}}$ & Threshold for signal demodulation \\ \hline
$P_{\mathrm{e}}$ & Symbol error rate \\ \hline
\end{tabular}
\end{table}

\section{Photon-Counting Statistics in Receivers}
\subsection{Poisson Arrival Process}
Photon arrival processes in photon-limited optical links are commonly modeled as Poisson processes, as weak light photons obey Poisson statistics \cite{ref33}. Consider a stochastic point process $N$ representing photon arrivals at time instants $\left\{ {{t_i}} \right\},i = 0,1,2, \cdots$. Let $N\left( t \right)$ denote the random variable for the number of arrivals within time interval $\left( {0,t} \right)$. For weak light sources, the photon arrival process follows Poisson distribution ${N_{\rm{s}}}\left( t \right) \!\sim\! \pi \left[ {\left( {{\lambda _{\rm{s}}} + {\lambda _{\rm{b}}}} \right)t} \right]$, where ${\lambda _{\rm{s}}},{\lambda _{\rm{b}}}$ are the average incident signal and background photon rates, respectively. For an ideal single-photon detector (ignoring dead time) with photon detection efficiency (PDE) ${p_{\rm{d}}}$, each incident photon is detected independently with probability ${p_{\rm{d}}}$. Dark carriers, generated at average rate ${\lambda _{\rm{d}}}$, also contribute to counts. The detected photon process remains Poisson-distributed: $N\left( t \right) \!\sim\! \pi \left( {\lambda t} \right)$, where $\lambda  \triangleq {p_{\rm{d}}}\left( {{\lambda _{\rm{s}}} + {\lambda _{\rm{b}}}} \right) + {\lambda _{\rm{d}}}$. Hence, the PMF of ideal single-photon detector detecting $k$ photon counts in time interval  is \cite{ref34}:
\begin{equation}\label{eq1}
P\left\{ {N\left( t \right) = k} \right\} = \frac{{{{\left( {\lambda t} \right)}^k}}}{{k!}}{{\rm{e}}^{ - \lambda t}}
\end{equation}
\textit{Proof}: Please refer to Appendix A. \\ where $\lambda  \ge 0$ represents the average detected carrier rate.

Furthermore, the Poisson nature of the photon arrival process implies that the time instant at which a SPAD generates an avalanche event (i.e., registers a photon counts) follows an exponential distribution. The probability density function (PDF) for the time $t$ of the first detected photon counts is \cite{ref34}:
\begin{equation}\label{eq2}
   f\left( t \right) = \lambda {{\rm{e}}^{ - \lambda t}},t \ge 0
\end{equation}

Beyond the desired optical signal photons, carriers generated by background radiation and dark counts (thermal noise, afterpulse) contribute indistinguishable false photon counts. To mitigate the inherent randomness of single-photon detection, symbol information is recovered by integrating photon counts over the symbol duration. Following each successful detection, the SPAD enters a deterministic dead time. Consequently, the continuous optical pulse stream is converted into a sequence of discrete avalanche events. 

Dead time introduces nonlinear distortion in photon counts, particularly pronounced at medium and high optical intensities. Maintaining an appropriate dynamic range and optimal communication performance necessitates a maximum photon counts on the order of several dozen per symbol. The maximum photon counts ${k_{\max }}$ depends on the dead time ${\tau _{\rm{d}}}$, symbol duration ${T_{\rm{s}}}$ and array scale ${N_{{\rm{A}}}}$ (i.e. the number of SPAD  pixels in array). For this detection scheme, ${k_{\max }}$ is given by:
\begin{align}\label{eq3}
{k_{\max }} = {N_{{\rm{A}}}}\left\lceil {\frac{{{T_{\rm{s}}}}}{{{\tau _{\rm{d}}}}}} \right\rceil
\end{align}
where $\left\lceil  \cdot  \right\rceil $denotes the ceiling function.

We then define dead time ratio to characterize the system timing constraint as: 
\begin{align}\label{eq72}
\xi  = \frac{\tau _{\rm{d}}}{T_{\rm{s}}}
\end{align}
For a single SPAD receiver, distinguishing symbol information via photon counts requires $\xi  < 1$ to maintain ${k_{\max }} > 1$. Due to the dead time, only a portion of photons are detected. To make decision, we must know the prior information of the waveform of laser pulse. To facilitate theoretical analysis, the transmitted laser pulse is assumed to be an ideal square wave.

To increase the maximum photon counts without extending symbol duration (i.e., decreasing data rate), SPADs are implemented as pixel arrays. Compared to single SPAD, SPAD array processes more avalanche events within symbol duration. In this case, the symbol duration can be equal to or even shorter than dead time ($\xi \ge 1$) and the error performance will outperform the single SPAD. When $\xi \ge 1$, a high-speed communication system is realized, which can break through the limitation of the dead time.

\subsection{Low- and Medium-Speed Systems ($\xi  < 1$)}

\begin{figure}[!t]
\centering
\includegraphics[width=3.3in]{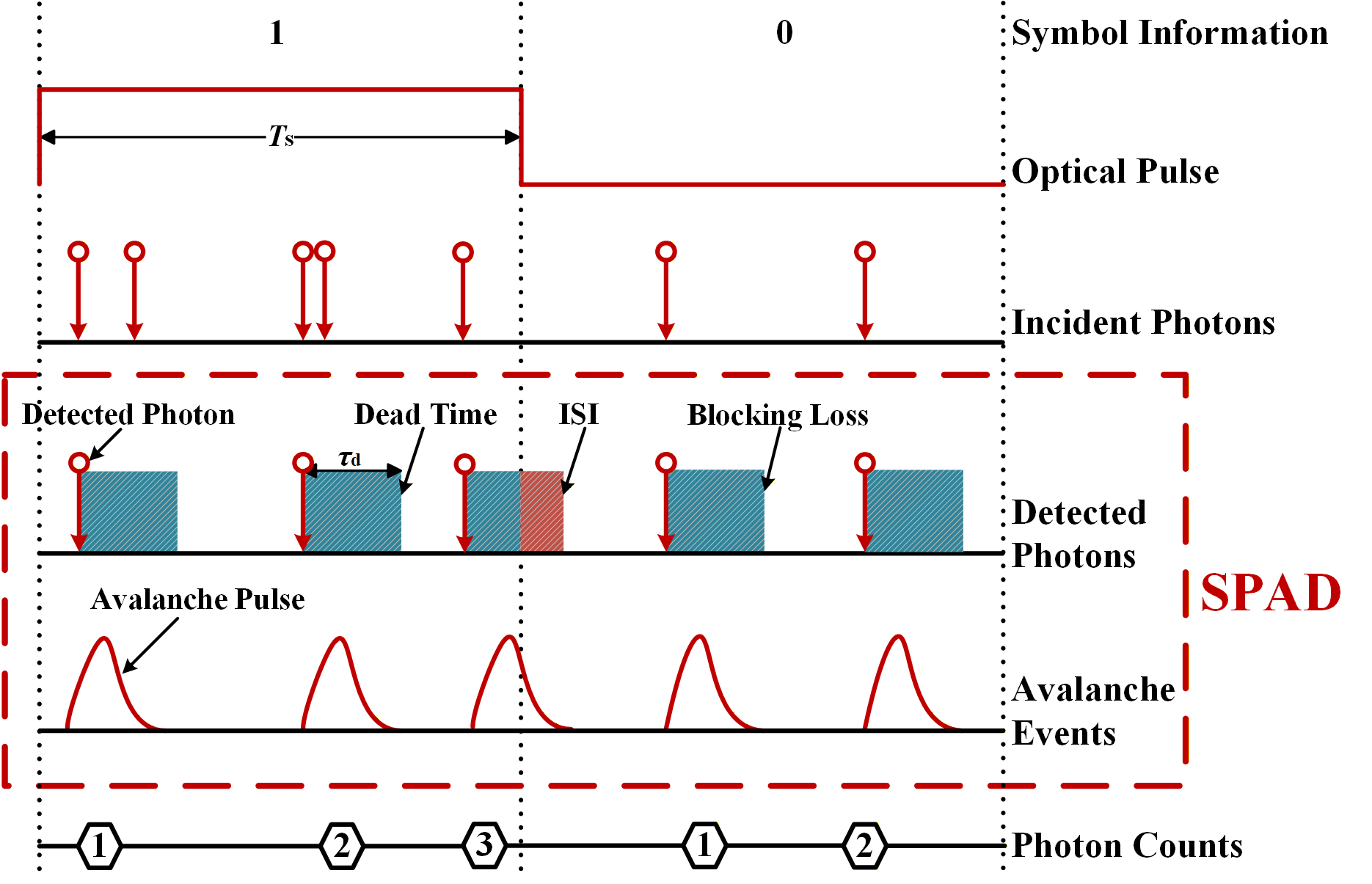}
\caption{Detection scheme of single SPAD-based photon-counting receiver for low- and medium-speed systems ($\xi = 0.25$).}
\label{fig_1}
\end{figure}

We first derive the PMF of photon counts for the regime $\xi  < 1$, where the symbol duration is longer than dead time. The detection scheme is illustrated in Fig. \ref{fig_1}. As illustrated in Fig. \ref{fig_1}, when the last photon counts occurs in time interval $\left( {{T_{\rm{s}}} - {\tau _{\rm{d}}},{T_{\rm{s}}}} \right)$, dead time persists into next symbol, causing ISI. This ISI reduces the effective integration time for symbol duration, thereby decreasing its photon counts. Critically, ISI in the current symbol is determined solely by the last photon arrival time in the preceding symbol. This dead-time-induced ISI introduces memory into the photon-counting channel. Let the stochastic variable ${T_{\rm{L}}}$ denote the arrival time of the last photon in symbol duration. We define $
p\left( {k,t} \right)$ as the probability that $k$ photons have been detected during time interval $\left( {0,t} \right)$. The PDF of ${T_{\rm{L}}}$ is derived as:
\begin{equation}\label{eq4}
{f_{{T_{\rm{L}}}}}\left( t \right) = \lambda {{\rm{e}}^{ - \lambda \left( {{T_{\rm{s}}} - t} \right)}},0 \le t \le {T_{\rm{s}}}
\end{equation}
\textit{Proof}: Please refer to Appendix B.

ISI occurs only when ${T_{\rm{s}}} - {\tau _{\rm{d}}} \le {T_{\rm{L}}} \le {T_{\rm{s}}}$. We define ${T_{{\rm{ISI}}}} = {T_{\rm{L}}} + {\tau _{\rm{d}}} - {T_{\rm{s}}}$ as the residual dead time from the previous symbol. Using (\ref{eq4}), the PMF of ${T_{{\rm{ISI}}}}$ can be obtained:
\begin{dmath}\label{eq5}
{f_{{T_{{\rm{ISI}}}}}}\left( t \right) = {f_{{T_{\rm{L}}}}}\left( {t - {\tau _{\rm{d}}} + {T_{\rm{s}}}} \right) + P\left\{ {{T_{{\rm{ISI}}}} = 0} \right\}\delta \left( t \right) \\
 = \lambda {{\rm{e}}^{ - \lambda \left( {{\tau _{\rm{d}}} - t} \right)}}\left[ {u\left( t \right) - u\left( {t - {\tau _{\rm{d}}}} \right)} \right] + \underbrace {P\left\{ {{T_{{\rm{ISI}}}} = 0} \right\}}_{ = 1 - \int_0^{{\tau _{\rm{d}}}} {{f_{{T_{\rm{L}}}}}\left( t \right)dt} }\delta \left( t \right) \\
= \lambda {{\rm{e}}^{ - \lambda \left( {{\tau _{\rm{d}}} - t} \right)}}\left[ {u\left( t \right) - u\left( {t - {\tau _{\rm{d}}}} \right)} \right] + {{\rm{e}}^{ - \lambda {\tau _{\rm{d}}}}}\delta \left( t \right) 
\end{dmath}

For the case $\xi  < 1$, the photon arrival process is modeled as a renewal process. Let ${f_{{T_1},{T_2}, \cdots ,{T_k}}}\left( {{t_1},{t_2}, \cdots ,{t_k}} \right)$ represents the joint PDF that the first photon counts occurs in the interval between $\left( {{t_1},{t_1} + d{t_1}} \right)$, the second photon counts between $\left( {{t_2},{t_2} + d{t_2}} \right)$, …, and the $k$-th photon counts between $\left( {{t_k},{t_k} + d{t_k}} \right)$. Provided that ${t_1},{t_2}, \cdots ,{t_k}$ are ordered ${t_1} < {t_2} <  \cdots  < {t_k}$. For Poisson processes with dead time, the $k$-order product density relates to first-order densities \cite{ref35}:
\begin{dmath}\label{eq6}
{f_{{T_1},\!\cdots\!,{T_k}}}\left( {{t_1},\!\cdots\!,{t_k}} \right)\!=\!{f_1}\left( {{t_1}} \right){f_2}\left( {t_2\!-\!t_1} \right)\!\cdots\!{f_2}\left( {t_k\!-\!t_{k - 1}} \right)
\end{dmath}

Ignoring ISI, the waiting time until the first photon ${f_1}\left( t \right)$ follows an exponential distribution. Subsequent inter-arrival times ${f_2}\left( t \right)$ follow a dead-time-modified exponential distribution. The PDFs are expressed as:
\begin{align}
{f_1}\left( t \right) &= \lambda {{\rm{e}}^{ - \lambda t}}u\left( t \right) \label{eq7}\\ {f_2}\left( t \right) &= \lambda {{\rm{e}}^{ - \lambda \left( {t - {\tau _{\rm{d}}}} \right)}}u\left( {t - {\tau _{\rm{d}}}} \right) \label{eq8} 
\end{align}

Hence, leveraging the independent increments property of the underlying Poisson process, the probability of detecting $k$-th photon counts at time instant $t$ without ISI is \cite{ref35}
\begin{align}\label{eq9}
&f_{{\rm{NISI}}}^{\rm{L}}\left( {k,t} \right)\!=\!\int_0^t { \cdots \int_0^t {{f_{{T_1},\!\cdots\!,{T_k}}}\left( {{t_1},\!\cdots\!,{t_k}} \right)d{t_1}\!\cdots\!d{t_k}}}  \\
& = \int_0^t {d{t_k}\int_0^{{t_k}} {d{t_{k - 1}} \cdots \int_0^{{t_2}} {{f_1}\left( {{t_1}} \right){f_2}\left( {{t_2} - {t_1}} \right) \cdots {f_2}\left( {{t_k} - {t_{k - 1}}} \right)d{t_1}}}} \nonumber
\end{align}

Based on renewal process theory, the probability of detecting $k$ photon counts in time interval $\left( {0,t} \right)$ without ISI, $p_{{\rm{NISI}}}^{\rm{L}}\left( {k,t} \right)$, can be written as follows:
\begin{equation}\label{eq10}
p_{{\rm{NISI}}}^{\rm{L}}\left( {k,t} \right) = \int_0^t {\left[ {f_{{\rm{NISI}}}^{\rm{L}}\left( {k,t'} \right) - f_{{\rm{NISI}}}^{\rm{L}}\left( {k + 1,t'} \right)} \right]dt'}  
\end{equation}

This aligns with established models for single SPAD receivers ignoring ISI \cite{ref28}:
\begin{dmath}\label{eq11}
p_{{\rm{NISI}}}^{\rm{L}}\left( {k,t} \right) = 
\begin{cases}
{\sum\limits_{i = 0}^k {\psi \left( {i,\lambda \left[ {t - k{\tau _{\rm{d}}}} \right]} \right)}  - \sum\limits_{i = 0}^{k - 1} {\psi \left( {i,\lambda \left[ {t - \left( {k - 1} \right){\tau _{\rm{d}}}} \right]} \right)}} & ,k < \left\lceil \frac{T_{\rm{s}}}{\tau_{\rm{d}}} \right\rceil \\
{1 - \sum\limits_{i = 0}^{k - 1} {\psi \left( {i,\lambda \left[ {t - \left( {k - 1} \right){\tau _{\rm{d}}}} \right]} \right)} } & ,k = \left\lceil \frac{T_{\rm{s}}}{\tau_{\rm{d}}} \right\rceil
\end{cases}
\end{dmath}
where $\psi \left( {i,\lambda } \right) = \frac{{{\lambda ^i}}}{{i!}}{{\rm{e}}^{ - \lambda }}$, $\left\lceil {{{{T_{\rm{s}}}} \mathord{\left/ {\vphantom {{{T_{\rm{s}}}} {{\tau _{\rm{d}}}}}} \right.\kern-\nulldelimiterspace} {{\tau _{\rm{d}}}}}} \right\rceil$ denotes the maximum photon counts per SPAD pixel per symbol.

We now extend this framework to model photon-counting statistics impaired by ISI. Crucially, photon counts in time interval $\left( {0,t} \right)$ depend not only on the number of detected photons, but also on the residual dead time propagated from the preceding symbol. Based on (\ref{eq9}), the full expression of detecting $k$ photon counts in time interval $\left( {0,t} \right)$ with ISI, $p_{{\rm{ISI}}}^{\rm{L}}\left( {k,t} \right)$, can be obtained by marginalizing over the distribution of :
\begin{dmath}\label{eq12}
p_{{\rm{ISI}}}^{\rm{L}}\left( {k,t} \right) = \int_0^{{\tau _{\rm{d}}}} {p_{{\rm{NISI}}}^{\rm{L}}\left( {t''} \right){f_{{T_{{\rm{ISI}}}}}}\left( {t - t''} \right)dt''} \\
 = \int_0^{{\tau _{\rm{d}}}} {\left\{ {\int_0^{t''} {\left[ {f_{{\rm{NISI}}}^{\rm{L}}\left( {k,t'} \right) - f_{{\rm{NISI}}}^{\rm{L}}\left( {k + 1,t'} \right)} \right]dt'} } \right\}{f_{{T_{{\rm{ISI}}}}}}\left( {t - t''} \right)dt''} \\
= {f_{{T_{{\rm{ISI}}}}}}\left( t \right) \otimes \int_0^t {\left[ {f_{{\rm{NISI}}}^{\rm{L}}\left( {k,t'} \right) - f_{{\rm{NISI}}}^{\rm{L}}\left( {k + 1,t'} \right)} \right]dt'} 
\end{dmath}
where $\otimes$ is convolution operator. Substituting (\ref{eq5}), (\ref{eq7}), and (\ref{eq8}) into (\ref{eq12}), yields the closed-form solution (\ref{eq13}) shown at the top of next page.

\begin{figure*}[!t]
\begin{dmath}\label{eq13}
 p_{{\rm{ISI}}}^{\rm{L}}\left( {k,t} \right) = 
\begin{cases}
 \hspace{1em} \left\langle {\sum\limits_{i = 0}^k {\frac{{{\lambda ^i}}}{{i!}}{{\left[ {t - k{\tau _{\rm{d}}}} \right]}^i}{{\rm{e}}^{ - \lambda \left[ {t - \left( {k - 1} \right){\tau _{\rm{d}}}} \right]}}}  - \sum\limits_{i = 0}^{k - 1} {\frac{{{\lambda ^i}}}{{i!}}{{\left[ {t - \left( {k - 1} \right){\tau _{\rm{d}}}} \right]}^i}{{\rm{e}}^{ - \lambda \left[ {t - \left( {k - 2} \right){\tau _{\rm{d}}}} \right]}}} } \right\rangle \\
 +  \left\langle {\sum\limits_{i = 0}^{k - 1} {\left( {1 - \frac{1}{{{2^{k - i}}}}} \right)\frac{{{\lambda ^i}}}{{i!}}{{\left[ {t - \left( {k - 1} \right){\tau _{\rm{d}}}} \right]}^i}{{\rm{e}}^{ - \lambda \left[ {t - \left( {k - 2} \right){\tau _{\rm{d}}}} \right]}} - \sum\limits_{i = 0}^{k - 1} {\left( {1 - \frac{1}{{{2^{k - i}}}}} \right)\frac{{{\lambda ^i}}}{{i!}}{{\left[ {t - k{\tau _{\rm{d}}}} \right]}^i}{{\rm{e}}^{ - \lambda \left[ {t - k{\tau _{\rm{d}}}} \right]}}} } }\right\rangle \\
 + \left\langle {\sum\limits_{i = 0}^k {\left( {1 - \frac{1}{{{2^{k - i + 1}}}}} \right)\frac{{{\lambda ^i}}}{{i!}}{{\left[ {t - \left( {k + 1} \right){\tau _{\rm{d}}}} \right]}^i}{{\rm{e}}^{ - \lambda \left[ {t - \left( {k + 1} \right){\tau _{\rm{d}}}} \right]}}}  - \sum\limits_{i = 0}^k {\left( {1 - \frac{1}{{{2^{k - i + 1}}}}} \right)\frac{{{\lambda ^i}}}{{i!}}{{\left[ {t - k{\tau _{\rm{d}}}} \right]}^i}{{\rm{e}}^{ - \lambda \left[ {t - \left( {k - 1} \right){\tau _{\rm{d}}}} \right]}}} } \right\rangle & ,k < \left\lceil \frac{T_{\rm{s}}}{\tau_{\rm{d}}} \right\rceil \\
\hspace{1em} {{{\rm{e}}^{ - \lambda {\tau _{\rm{d}}}}} - \sum\limits_{i = 0}^{k - 1} {\frac{{{\lambda ^i}}}{{i!}}{{\left[ {t - \left( {k - 1} \right){\tau _{\rm{d}}}} \right]}^i}{{\rm{e}}^{ - \lambda \left[ {t - \left( {k - 2} \right){\tau _{\rm{d}}}} \right]}}}} & ,k = \left\lceil \frac{T_{\rm{s}}}{\tau_{\rm{d}}} \right\rceil
\end{cases}
\end{dmath}
\textit{Proof}: Please refer to Appendix C.
\vspace{1em}
\hrulefill 
\end{figure*}

Regarding (\ref{eq13}), the first two terms represent the PMF without ISI. The remaining four terms model the PMF corrections for ISI-induced distortion, specifically accounting for the reduction in effective symbol duration. To derive a tractable closed-form solution, we assume the preceding symbol information identical to the current symbol (i.e., equal optical power). While this simplification facilitates computation, it introduces estimation bias: Overestimation occurs for high-power symbols (where the modeled ISI duration exceeds the actual ISI); Underestimation occurs for low-power symbols (where the modeled ISI duration is less than the actual ISI). To obtain an accurate PMF, we incorporate the mean of $p_{{\rm{ISI}}}^{\rm{L}}$ and $p_{{\rm{NISI}}}^{\rm{L}}$:
\begin{equation}\label{eq14}
{p^{\rm{L}}}\left( {k,t} \right) = \frac{1}{2}\left[ {p_{{\rm{ISI}}}^{\rm{L}}\left( {k,t} \right) + p_{{\rm{NISI}}}^{\rm{L}}\left( {k,t} \right)} \right]   
\end{equation}

For an SPAD array comprising ${N_{{\rm{A}}}}$ pixels, the total photon counts are the summation of counts from all constituent pixels. Modeling each pixel's output as independent, the PMF for the SPAD array follows a multinomial distribution:
\begin{dmath}\label{eq15}
p_{{\rm{A}}}^{\rm{L}}\left( {k,t} \right)\!=\!\sum\limits_{ \begin{array}{*{20}{c}} \scriptstyle\sum\nolimits_{i = 0}^{\left\lceil \frac{T_{\rm{s}}}{\tau_{\rm{d}}} \right\rceil} {{k_i}}\!=\!{N_{{\rm{A}}}} \\
\scriptstyle\sum\nolimits_{i = 0}^{\left\lceil \frac{T_{\rm{s}}}{\tau_{\rm{d}}} \right\rceil } {i{k_i}}  = k \end{array}} \binom{{N_{{\rm{A}}}}}{{k_0},\!\cdots\!,{k_{\left\lceil \frac{T_{\rm{s}}}{\tau_{\rm{d}}} \right\rceil}}}
 \left\{ \begin{array}{l} {p^{\rm{L}}}{{\left( {0,t} \right)}^{{k_0}}}{p^{\rm{L}}}{{\left( {1,t} \right)}^{{k_1}}}  \\
\!\cdots\!{p^{\rm{L}}}{{\left( {\left\lceil \frac{T_{\rm{s}}}{\tau_{\rm{d}}} \right\rceil ,t} \right)}^{{k_{\left\lceil \frac{T_{\rm{s}}}{\tau_{\rm{d}}} \right\rceil}}}} \end{array} \right\}
\end{dmath}
where ${k_i}$ is the number of pixels, which output exactly $i$ photon counts. The multinomial coefficient is \cite{ref36}:
\begin{equation}\label{eq16}
 \binom{{N_{{\rm{A}}}}}{{k_0}, \cdots, {k_{\left\lceil \frac{T_{\rm{s}}}{\tau_{\rm{d}}} \right\rceil}}}
  = \frac{{{N_{{\rm{A}}}}!}}{{{k_0}!{k_1}! \cdots {k_{\left\lceil \frac{T_{\rm{s}}}{\tau_{\rm{d}}} \right\rceil}}!}}    
\end{equation}

\subsection{High-Speed Systems ($\xi \ge 1$)}

\begin{figure}[!t]
\centering
\includegraphics[width=3.3in]{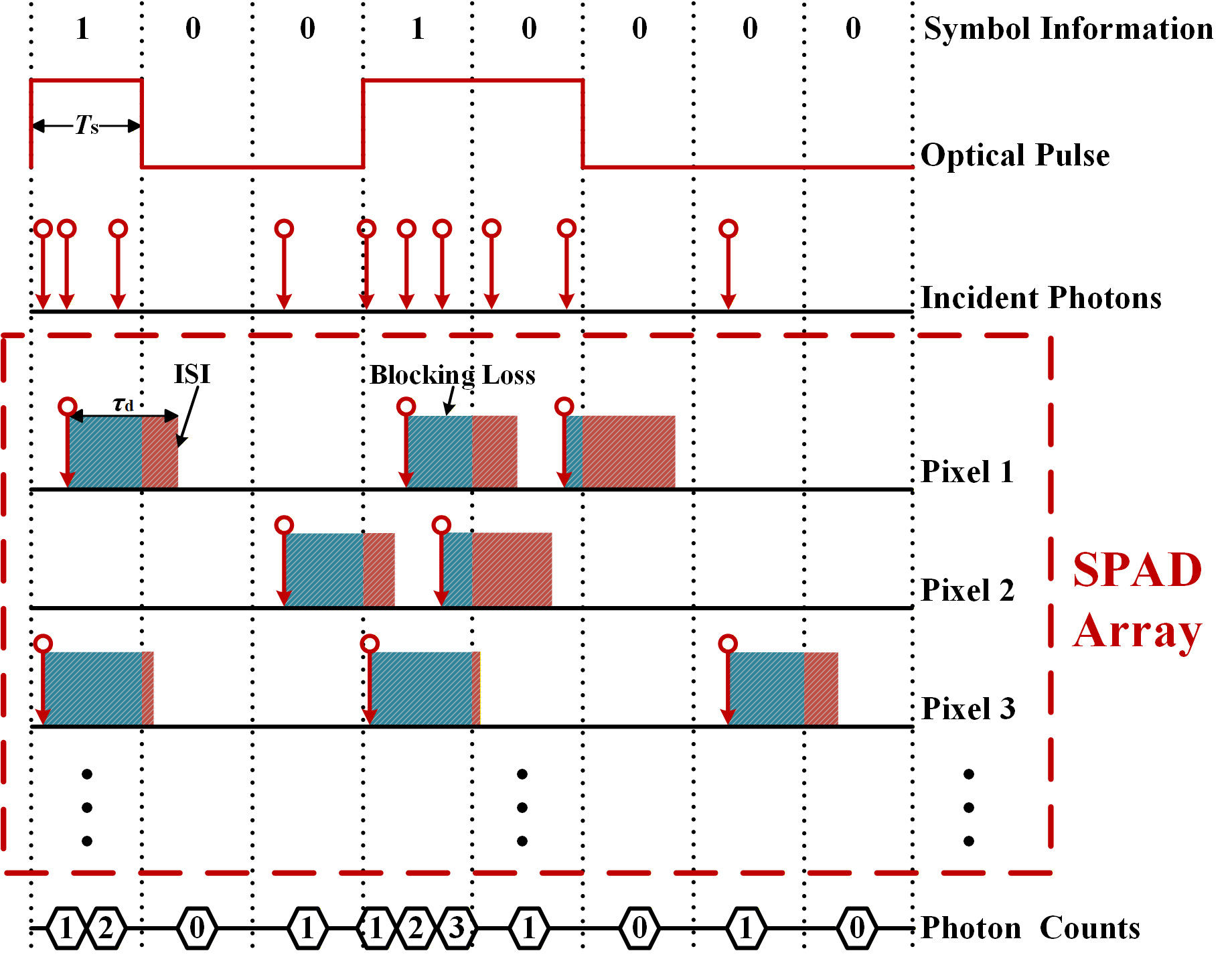}
\caption{Detection scheme of SPAD array-based photon-counting receiver for high-speed systems ($\xi  = 1$).}
\label{fig_2}
\end{figure}

For high-speed systems ($\xi \ge 1$), symbol duration is shorter than or equal to dead time. The detection scheme is presented in Fig. \ref{fig_2}, which permits at most one photon detection per pixel per symbol due to dead time constraints. Crucially, dead time persists across multiple symbol durations, causing ISI that invalidates renewal process modeling. Consequently, a Markov process framework is required to derive the PMF of photon counts.
We first consider the case $\xi  = 1$(i.e. dead time is equal to symbol duration). For the initial symbol (no ISI), the PDF of the photon arrival time simplifies from (\ref{eq2}):
\begin{equation}\label{eq17}
f_{{T_{\rm{L}}}}^1\left( t \right) = \lambda {{\rm{e}}^{ - \lambda t}},0 \le t \le {T_{\rm{s}}} 
\end{equation}

For subsequent symbols, define ${T_{{\rm{ISI}}}} = {T_{\rm{L}}}$ as the residual dead time from the previous symbol. The PDF of ${T_{{\rm{ISI}}}}$ derived from (\ref{eq17}), is:
\begin{dmath}\label{eq18}
f_{{T_{{\rm{ISI}}}}}^1\!\left( t \right)\! =\! {f_{{T_{\rm{L}}}}}\left( {t - {\tau _{\rm{d}}} + {T_{\rm{s}}}} \right) + P\left\{ {{T_{{\rm{ISI}}}} 
 = 0} \right\}\delta \left( t \right) \\
\! = \!\lambda {{\rm{e}}^{\! -\! \lambda t}}\left[ {u\left( t \right) \!- u\left( {t\! -\! {T_{\rm{s}}}} \right)} \right]\! +\! \underbrace {P\left\{ {{T_{{\rm{ISI}}}}\! =\! 0} \right\}}_{ = 1 - \int_0^{{T_{\rm{s}}}} {f_{{T_{\rm{L}}}}^1\left( t \right)dt} }\!\delta \left( t \right)\\
\! =\! \lambda {{\rm{e}}^{ - \lambda t}}\left[ {u\left( t \right) - u\left( {t - {T_{\rm{s}}}} \right)} \right] + {{\rm{e}}^{ - \lambda {T_{\rm{s}}}}}\delta \left( t \right)  
\end{dmath}

ISI propagates sequentially between symbols. The PDF transition function between consecutive symbols is:
\begin{align}\label{eq19}
f_{T_{\rm{ISI}}}^n\left( t \right)\!=\!\int_0^{{T_{\rm{s}}}} {\left\{ \begin{array}{l}
\lambda {{\rm{e}}^{ \!-\! \lambda \left( {t \!-\! t'} \right)}}\left[ {u\left( {t - t'} \right)\! -\! u\left( {t \!-\! t' \!- \!{T_{\rm{s}}}} \right)} \right]\\
 \!+\! {{\rm{e}}^{ - \lambda \left( {{T_{\rm{s}}} - t'} \right)}}\delta \left( t \right)
\end{array} \right\}\!f_{T_{\rm{ISI}}}^{n \!- \!1}\!\left( {t'} \right)dt'} 
\end{align}

Iterative substitution of (\ref{eq18}) into (\ref{eq19}) yields the PDF of ${T_{{\rm{ISI}}}}$ for the $n$-th symbol. While no closed-form solution exists, numerical convergence occurs by $n \ge 5$. Thus, we approximate the steady-state PDF using the sixth symbol, which is presented as (\ref{eq20}) at the top of this page.
\begin{figure*}[!t]
\begin{align}\label{eq20}
{\tilde f_{{T_{{\rm{ISI}}}}}}\left( t \right) = \lambda {{\rm{e}}^{ - \lambda t}}\left\{ \begin{array}{l}
 \frac{{2{{\left( {\lambda t} \right)}^4} + 4\left( {{\lambda ^4}{t^3}{T_{\rm{s}}} + {\lambda ^4}t{T_{\rm{s}}}^3} \right) + {{\left( {\lambda {T_{\rm{s}}}} \right)}^4}}}{{24}}{{\rm{e}}^{ - \lambda {T_{\rm{s}}}}}\\
 + \frac{{{{\left( {\lambda t} \right)}^3} + 6{\lambda ^3}{t^2}{T_{\rm{s}}} + 12{\lambda ^3}t{T_{\rm{s}}}^2 + 6{{\left( {\lambda {T_{\rm{s}}}} \right)}^3}}}{6}{{\rm{e}}^{ - 2\lambda {T_{\rm{s}}}}}\\
 + \frac{{{{\left( {\lambda t} \right)}^2} + 6{\lambda ^2}t{T_{\rm{s}}} + 9{{\left( {\lambda {T_{\rm{s}}}} \right)}^2}}}{2}{{\rm{e}}^{ - 3\lambda {T_{\rm{s}}}}}\\
 + \frac{{\lambda t + 4\lambda {T_{\rm{s}}}}}{2}{{\rm{e}}^{ - 4\lambda {T_{\rm{s}}}}} + {{\rm{e}}^{ - 5\lambda {T_{\rm{s}}}}}
 + \frac{{{{\left( {\lambda t} \right)}^5}}}{{120}} 
\end{array} \right\}  \left[  {u\left( t \right)  - u\left( {t - {T_{\rm{s}}}} \right)} \right]
+ \left\{ \begin{array}{l}
\frac{{{{\left( {\lambda {T_{\rm{s}}}} \right)}^5}}}{{120}}{{\rm{e}}^{ - \lambda {T_{\rm{s}}}}} \\ + \frac{{11{{\left( {\lambda {T_{\rm{s}}}} \right)}^4}}}{{24}}{{\rm{e}}^{ - 2\lambda {T_{\rm{s}}}}}\\
 + \frac{{25{{\left( {\lambda {T_{\rm{s}}}} \right)}^3}}}{6}{{\rm{e}}^{ - 3\lambda {T_{\rm{s}}}}}\\
 + 8{\left( {\lambda {T_{\rm{s}}}} \right)^2}{{\rm{e}}^{ - 4\lambda {T_{\rm{s}}}}}\\
 + 5\lambda {T_{\rm{s}}}{{\rm{e}}^{ - 5\lambda {T_{\rm{s}}}}}  \\ +{{\rm{e}}^{ - 6\lambda {T_{\rm{s}}}}}
\end{array} \right\}\delta \left( t \right)
\end{align}
\textit{Proof}: Please refer to Appendix D.
\vspace{1em}
\hrulefill 
\end{figure*}

\begin{figure}[!t]
\centering
\includegraphics[width=3.3in]{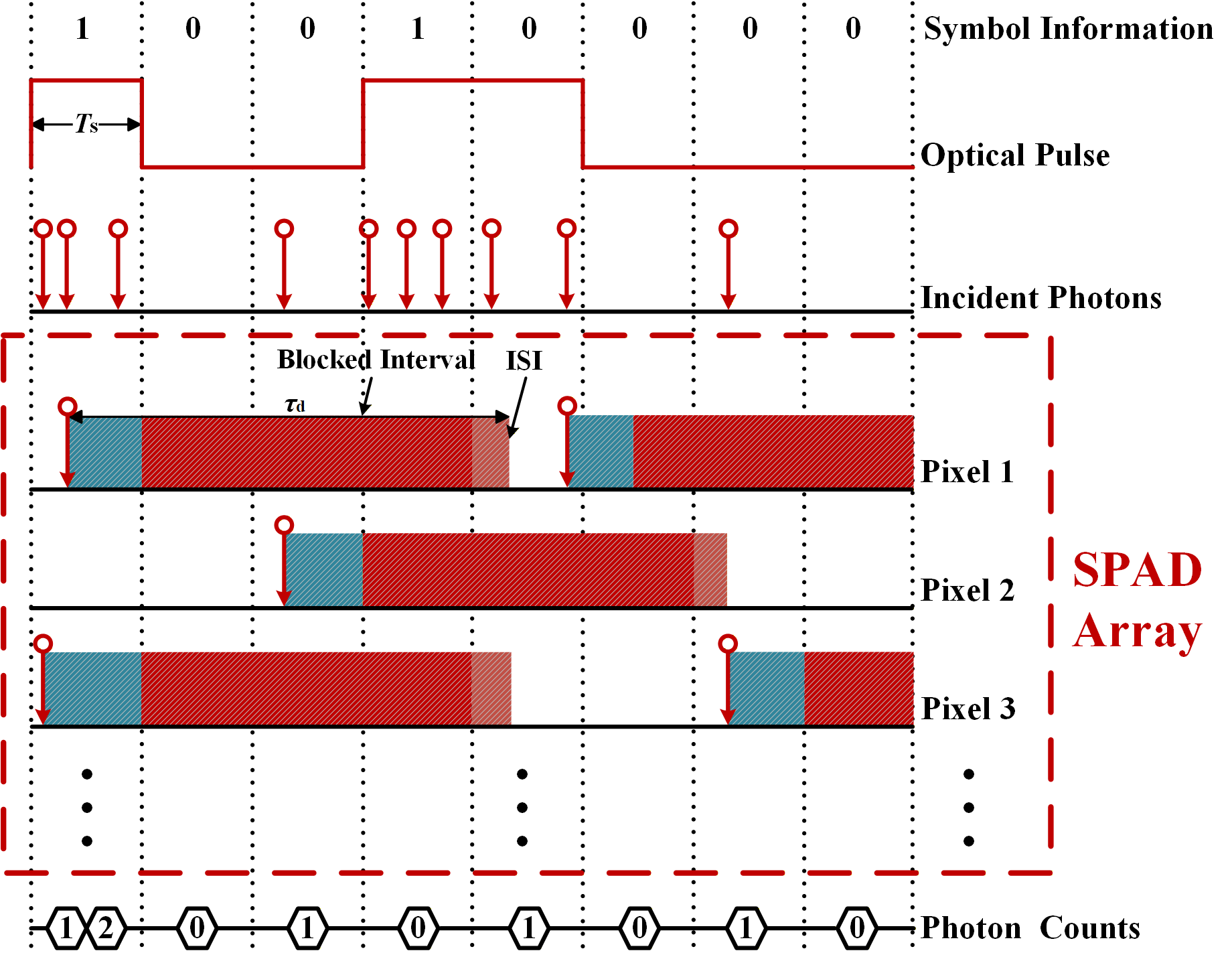}
\caption{Detection scheme of SPAD array-based photon-counting receiver for high-speed systems ($\xi  = 4$).}
\label{fig_3}
\end{figure}

Define $p_{{\rm{ISI}}}^{\rm{H}}\left( {1,{T_{\rm{s}}}} \right)$ as the trigger probability of a single SPAD pixel of detecting one photon per symbol with ISI. Based on (\ref{eq19}), the trigger probability is calculated by:
\begin{dmath}\label{eq21}
p_{{\rm{ISI}}}^{\rm{H}}\left( {1,{T_{\rm{s}}}} \right) = P\left\{ {{T_{{\rm{ISI}}}} > 0} \right\}\\
 = 1 - P\left\{ {{T_{{\rm{ISI}}}} = 0} \right\}\\
 = 1 - \int_{{0^ - }}^{{0^ + }} {{{\tilde f}_{{T_{{\rm{ISI}}}}}}\left( t \right)d} t\\
 = 1 - {{\rm{e}}^{ - \lambda {T_{\rm{s}}}}}\left\{ \begin{array}{l}
\frac{{{{\left( {\lambda {T_{\rm{s}}}} \right)}^5}}}{{5!}} + \frac{{11{{\left( {\lambda {T_{\rm{s}}}} \right)}^4}}}{{24}}{{\rm{e}}^{ - \lambda {T_{\rm{s}}}}} \\ + \frac{{25{{\left( {\lambda {T_{\rm{s}}}} \right)}^3}}}{6}{{\rm{e}}^{ - 2\lambda {T_{\rm{s}}}}}
 \\ + 8{\left( {\lambda {T_{\rm{s}}}} \right)^2}{{\rm{e}}^{ - 3\lambda {T_{\rm{s}}}}} \\ + 5\lambda {T_{\rm{s}}}{{\rm{e}}^{ - 4\lambda {T_{\rm{s}}}}} + {{\rm{e}}^{ - 5\lambda {T_{\rm{s}}}}}
\end{array} \right\} 
\end{dmath}
The complementary probability of no photon detection is $p_{{\rm{ISI}}}^{\rm{H}}\left( {0,{T_{\rm{s}}}} \right) = 1 - p_{{\rm{ISI}}}^{\rm{H}}\left( {1,{T_{\rm{s}}}} \right)$.

The last six exponential term in (\ref{eq21}) accounts for ISI-induced reduction in effective symbol duration. To improve accuracy, we incorporate the mean residual dead time, yielding the simplified trigger probability. Following the approach for low-speed systems, the mean of $p_{{\rm{ISI}}}^{\rm{H}}\left( {1,{T_{\rm{s}}}} \right)$ and $p_{{\rm{NISI}}}^{\rm{H}}\left( {1,{T_{\rm{s}}}} \right)$ is adopted:
\begin{equation}\label{eq22}
{p^{\rm{H}}}\left( {1,{T_{\rm{s}}}} \right) = \frac{1}{2}\left[ {p_{{\rm{ISI}}}^{\rm{H}}\left( {1,{T_{\rm{s}}}} \right) + p_{{\rm{NISI}}}^{\rm{H}}\left( {1,{T_{\rm{s}}}} \right)} \right]
\end{equation}
where the trigger probability of ISI-free case is
\begin{dmath}\label{eq23}
p_{{\rm{NISI}}}^{\rm{H}}\left( {1,{T_{\rm{s}}}} \right) = P\left\{ {N\left( {{T_{\rm{s}}}} \right) > k} \right\}
 = 1 - P\left\{ {N\left( {{T_{\rm{s}}}} \right) = 0} \right\}
 = 1 - {{\rm{e}}^{ - \lambda {T_{\rm{s}}}}}
\end{dmath}

\begin{figure}[!t]
\centering
\subfloat[Symbol detection states for PAM]{\includegraphics[width=3.3in]{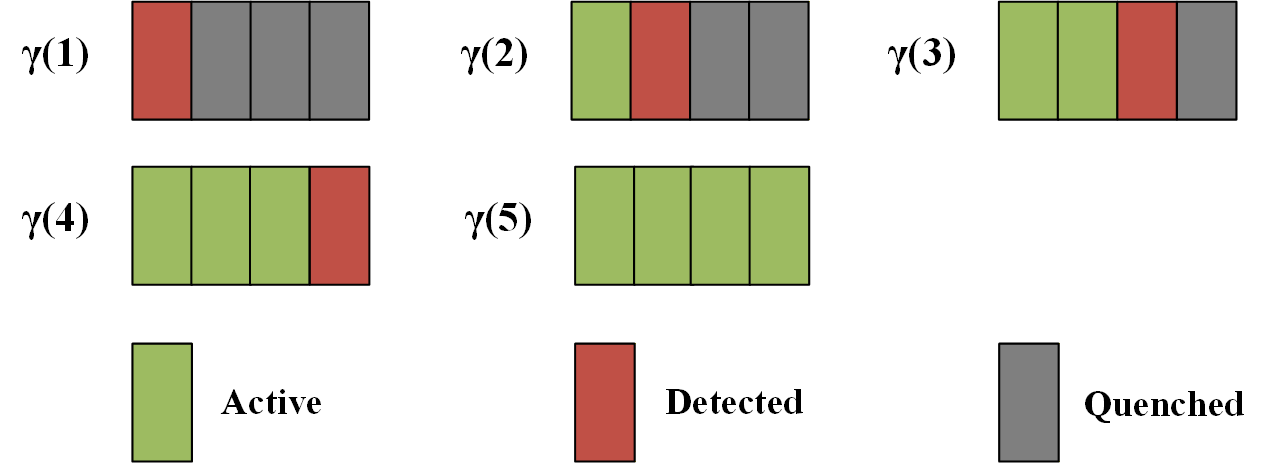}%
\label{fig_4_1}}
\hfil
\subfloat[States transition graph of Markov model]{\includegraphics[width=3.3in]{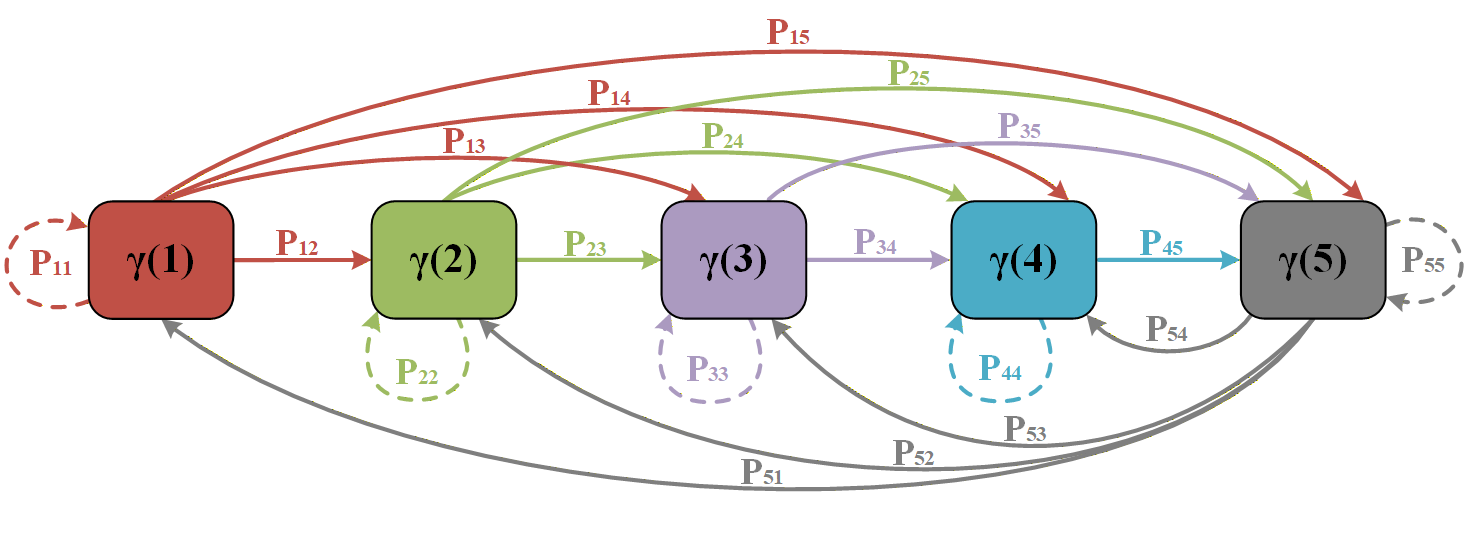}%
\label{fig_4_2}}
\caption{Symbol detection states for high-speed system ($\xi  = 4$).}
\label{fig_4}
\end{figure}

Then, we extend our analysis to the general case where the dead time ratio $\xi  \ge 1$ is a positive integer. Non-integer values yield prohibitively complex PMFs and are not considered here. As Fig. \ref{fig_3} illustrates, when dead time ratio is an integer $\xi  \ge 1$, the residual dead time spans exactly $\xi $ symbol durations, and the PMF of ${T_{{\rm{ISI}}}}$ mirrors the case $\xi  = 1$. At any symbol epoch, the SPAD occupies one of three states:

\textbf{(1) State 1 (Quenched):} SPAD is quenched by dead time.

\textbf{(2) State 2 (Detected):} SPAD is detecting a photon counts.

\textbf{(3) State 3 (Active):} SPAD is sensitive (capable of detecting one photon). 

Denote the steady-state probabilities of \textbf{State 1} as  $
{p_{{\rm{S}}}}\left( 0 \right)$ (SPAD is inactive). Denote $
{p_{{\rm{S}}}}\left( 1 \right)$ as the steady-state probabilities of \textbf{State 2} and \textbf{3} (SPAD is active).

We adopt a finite-state machine to model the photon detection behavior of an SPAD with respect to an integer dead time ratio ($\xi \ge 1$). The successive symbols are divided in to groups, where each group contains $\xi$ symbol durations. We define the probability of SPAD detecting photon counts at 1- to $\xi$-th symbol in the group as $\gamma \left( 1 \right),\gamma \left( 2 \right), \cdots ,\gamma \left( \xi  \right)$. In particularly, we define that $\gamma \left( {\xi  + 1} \right)$ represents the probability of detecting no photon counts in group. As Fig. \ref{fig_4}b depicted, the state of SPAD can be characterized by a Markov chain. The state evolution follows the transition matrix:
\begin{equation}\label{eq24}
     P_{\rm{T}} =
    \begin{bmatrix}
        \tilde{p}_1                & \tilde{p}_0 p_1          & \cdots & \tilde{p}_0 p_1 p_0^{\xi-2} & \tilde{p}_0 p_0^{\xi-1} \\
        0                          & \tilde{p}_1             & \cdots & \tilde{p}_0 p_1 p_0^{\xi-3} & \tilde{p}_0 p_0^{\xi-2} \\
        \vdots                     & \vdots                          & \ddots & \vdots                         & \vdots                         \\
        0                          & 0                               & \cdots & \tilde{p}_1                    & \tilde{p}_0                    \\
        p_1                        & p_1 p_0                   & \cdots & p_1 p_0^{\xi-1}             & p_0^{\xi}
    \end{bmatrix}
\end{equation}
where ${\tilde p_1} \triangleq {p^{\rm{H}}}\left( {1,{T_{\rm{s}}}} \right)$, ${\tilde p_0} \triangleq {p^{\rm{H}}}\left( {0,{T_{\rm{s}}}} \right)$ and ${p_1} \triangleq p_{{\rm{NISI}}}^{\rm{H}}\left( {1,{T_{\rm{s}}}} \right)$, ${p_0} \triangleq p_{{\rm{NISI}}}^{\rm{H}}\left( {0,{T_{\rm{s}}}} \right)$. For $M$-PAM signals, the optical intensity differs across symbols. The average trigger probability is therefore employed as ${\tilde p_1}{\rm{ = }}{\textstyle{\frac{1}{M}}}\sum\nolimits_{m = 1}^M {{p^{\rm{H}}} \left( {1,{T_{\rm{s}}},{\lambda _{\rm{s}}}\left( {{x_m}} \right)} \right)}$, ${\tilde p_0}{\rm{ = }}1 - {\tilde p_1}$ and ${p_1}{\rm{ = }}{\textstyle{\frac{1}{M}}}\sum\nolimits_{m = 1}^M {p_{{\rm{NISI}}}^{\rm{H}}\left( {1,{T_{\rm{s}}},{\lambda_{\rm{s}}}\left( {{x_m}} \right)} \right)} $, ${p_0} = 1 - {p_1}$. 

According to (\ref{eq23}) (\ref{eq24}), the probability of symbol states in the group $\gamma \left( 1 \right),\gamma \left( 2 \right), \cdots ,\gamma \left( {\xi  + 1} \right)$ can be uniquely solved by the Chapman–Kolmogorov equation as follows:
\begin{align}\label{eq25}
\!\left[ {\begin{array}{*{20}{c}}
{{{\tilde p}_1}}&0& \cdots &{{p_1}}\\
{{{\tilde p}_0}{p_1}}&{{{\tilde p}_1} }& \cdots &{{p_1}{p_0}}\\
 \vdots & \vdots  & \ddots & \vdots \\
{{{\tilde p}_0}{p_0}{}^{\xi\!-\!1}}&{{{\tilde p}_0}{p_0}{}^{\xi\!-\! 2}}& \cdots &{{p_0}{}^\xi  }\\
1&1& \cdots &1
\end{array}} \right] \! \left[ {\begin{array}{*{20}{c}}
\gamma (1)\\
\gamma (2)\\
 \vdots \\
\gamma (\xi)\\
\gamma (\xi\!+\!1)
\end{array}} \right]\!=\!\left[ {\begin{array}{*{20}{c}}
\gamma (1)\\
\gamma (2)\\
 \vdots \\
\gamma (\xi\!+\!1)\\
1
\end{array}} \right]  
\end{align}

The unique solution to linear equations is derived as follows:
\begin{align}\label{eq26}
\left\{ \begin{array}{l}
\gamma \left( 1 \right) = \gamma \left( 2 \right) =  \cdots  = \gamma \left( \xi  \right) = \dfrac{{{p_1}}}{{\xi {p_1} + {{\tilde p}_0}}}\\
\gamma \left( {\xi  + 1} \right) = \dfrac{{{{\tilde p}_0}}}{{\xi {p_1} + {{\tilde p}_0}}}
\end{array} \right.   
\end{align}
\textit{Proof}: Please refer to Appendix E.

Solving this linear system yields unique steady-state probabilities. Based on (\ref{eq26}), the probability of an arbitrary symbol being active is derived then:
\begin{align}\label{eq27}
{p_{{\rm{S}}}}\left( 1 \right) &= \sum\limits_{n = 1}^{\xi  + 1} {\gamma \left( n \right)P\left\{ {{\rm{act}}\left| {{\rm{D}} = n} \right.} \right\}} \\
 &= \sum\limits_{n = 1}^\xi  {\gamma \left( n \right)\left( {\frac{{n{\rm{ + }}1}}{{2\xi }}} \right)}  + \gamma \left( {\xi  + 1} \right)\left[ {\frac{{\left( {\xi  + 1} \right){p_1}{{\tilde p}_0} + 2{{\tilde p}_0}^2}}{{2\left( {\xi {p_1} + {{\tilde p}_0}} \right)}}} \right]\nonumber\\
 &= \frac{{\xi \left( {\xi  + 3} \right){p_1}^2 + \left( {3\xi  + 5} \right){p_1}{{\tilde p}_0}{\rm{ + 4}}{{\tilde p}_0}^2}}{{4{{\left( {\xi {p_1} + {{\tilde p}_0}} \right)}^2}}}\nonumber
\end{align}
where $P\left\{ {{\rm{act}}\left| {{\rm{D}} = n} \right.} \right\}$ is the condition probability of that a pixel is active given detection occurred in the $n$-th symbol or no detection in the group. It is expressed as:
\begin{align}
P\left\{ {{\rm{act}}\left| {{\rm{D}} = n \le \xi } \right.} \right\} &= \frac{1}{n}\sum\limits_{i = 1}^n {\frac{i}{\xi }}  = \frac{{n + 1}}{{2\xi }} \label{eq28} \\
P\left\{ {{\rm{act}}\left| {{\rm{D}} = \xi  + 1} \right.} \right\} &= \frac{1}{\xi }\sum\limits_{i = 1}^\xi  {\left( {\frac{i}{\xi }} \right)\left( {\sum\limits_{n = 1}^\xi  {\gamma \left( n \right)} } \right){\rm{ + }}\gamma \left( {\xi  + 1} \right)} \nonumber\\
 & = \frac{{1 + \xi }}{{2\xi }}\frac{{\xi {p_1}}}{{\xi {p_1} + {{\tilde p}_0}}} + \frac{{{{\tilde p}_0}}}{{\xi {p_1} + {{\tilde p}_0}}} \label{eq29} \\
 & = \frac{{\left( {\xi  + 1} \right){p_1}{{\tilde p}_0} + 2{{\tilde p}_0}^2}}{{2\left( {\xi {p_1} + {{\tilde p}_0}} \right)}}  \nonumber
\end{align}

As each pixel detects at most one photon per symbol, SPAD arrays are essential for signal recovery. The probability of $j$ SPAD pixels being active is:
\begin{align}\label{eq30}
{p_{{\rm{pix}}}}\left( j \right) = \binom{N_{\rm{A}}}{j} {p_{{\rm{S}}}}{\left( 1 \right)^j}{p_{{\rm{S}}}}{\left( 0 \right)^{{N_{{\rm{A}}}} - j}}    
\end{align}

Based on (\ref{eq30}), the PMF of photon counts can be derived:
\begin{align}\label{eq31}
&p_{{\rm{A}}}^{\rm{H}}\left( {k,{T_{\rm{s}}}} \right) = \sum\limits_{j = k}^{{N_{{\rm{A}}}}} {P\left\{ {{\rm{C}} = k\left| {{\rm{Pix}} = j} \right.} \right\}} {p_{{\rm{pix}}}}\left( j \right)\\
& = \sum\limits_{j = k}^{{N_{{\rm{A}}}}} {\binom{j}{k} {p^{\rm{H}}}{{\left( {1,{T_{\rm{s}}}} \right)}^k}{p^{\rm{H}}}{{\left( {0,{T_{\rm{s}}}} \right)}^{j - k}}} \binom{N_{\rm{A}}}{j} {p_{{\rm{S}}}}{\left( 1 \right)^j}{p_{{\rm{S}}}}{\left( 0 \right)^{{N_{{\rm{A}}}} - j}} \nonumber
\end{align}
where $P\left\{ {{\rm{C}} = k\left| {{\rm{Pix}} = j} \right.} \right\}$ is the condition probability of that detecting $k$ photon counts when $j$ SPAD pixels are active.

\subsection{PMF Comparative Analysis}
\begin{figure*}[!htb]
\centering
\subfloat[]{\includegraphics[width=3.3in]{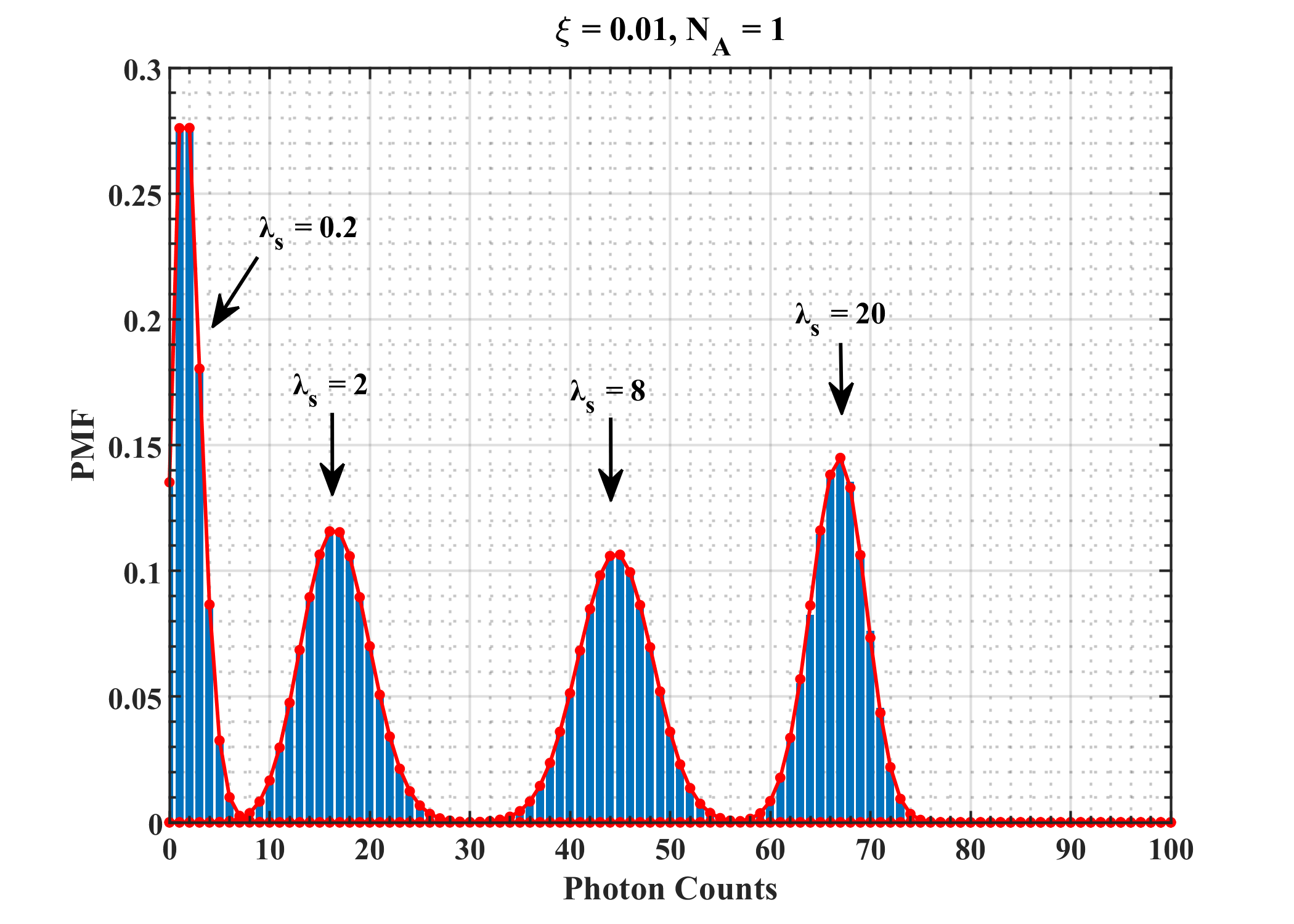}
\label{fig_5_1}}
\hfil
\subfloat[]{\includegraphics[width=3.3in]{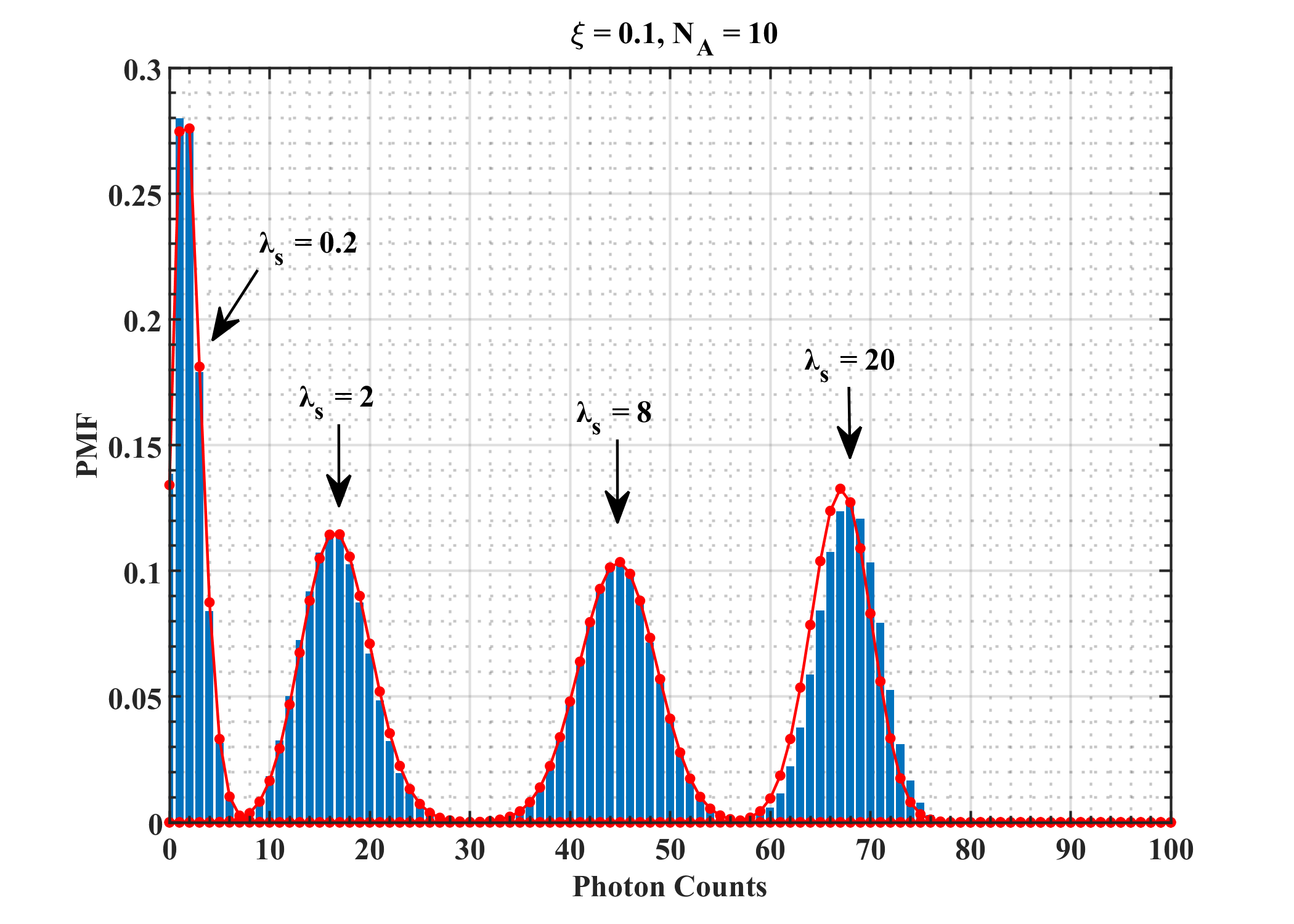}
\label{fig_5_2}}
\hfil
\subfloat[]{\includegraphics[width=3.3in]{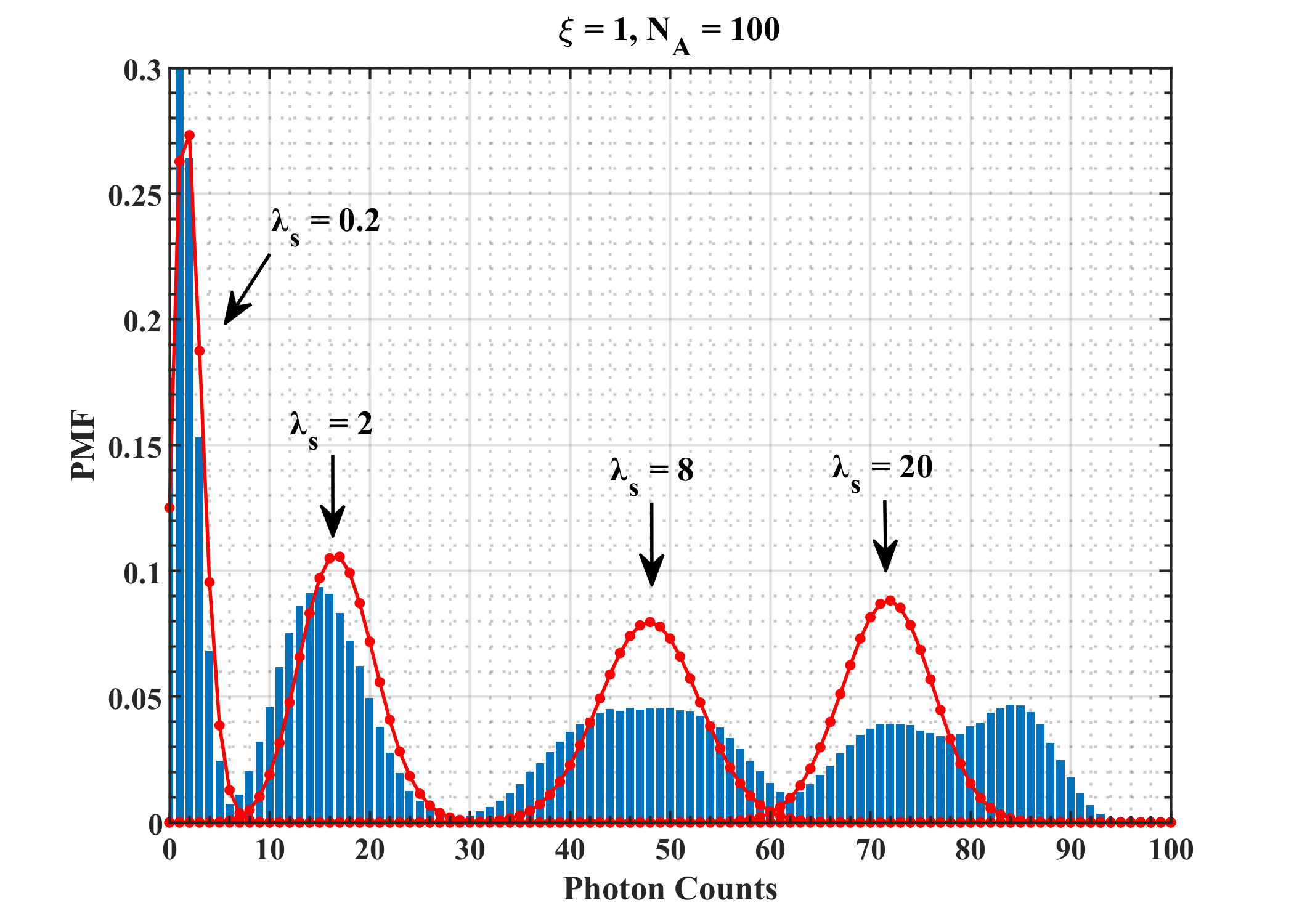}
\label{fig_5_3}}
\hfil
\subfloat[]{\includegraphics[width=3.3in]{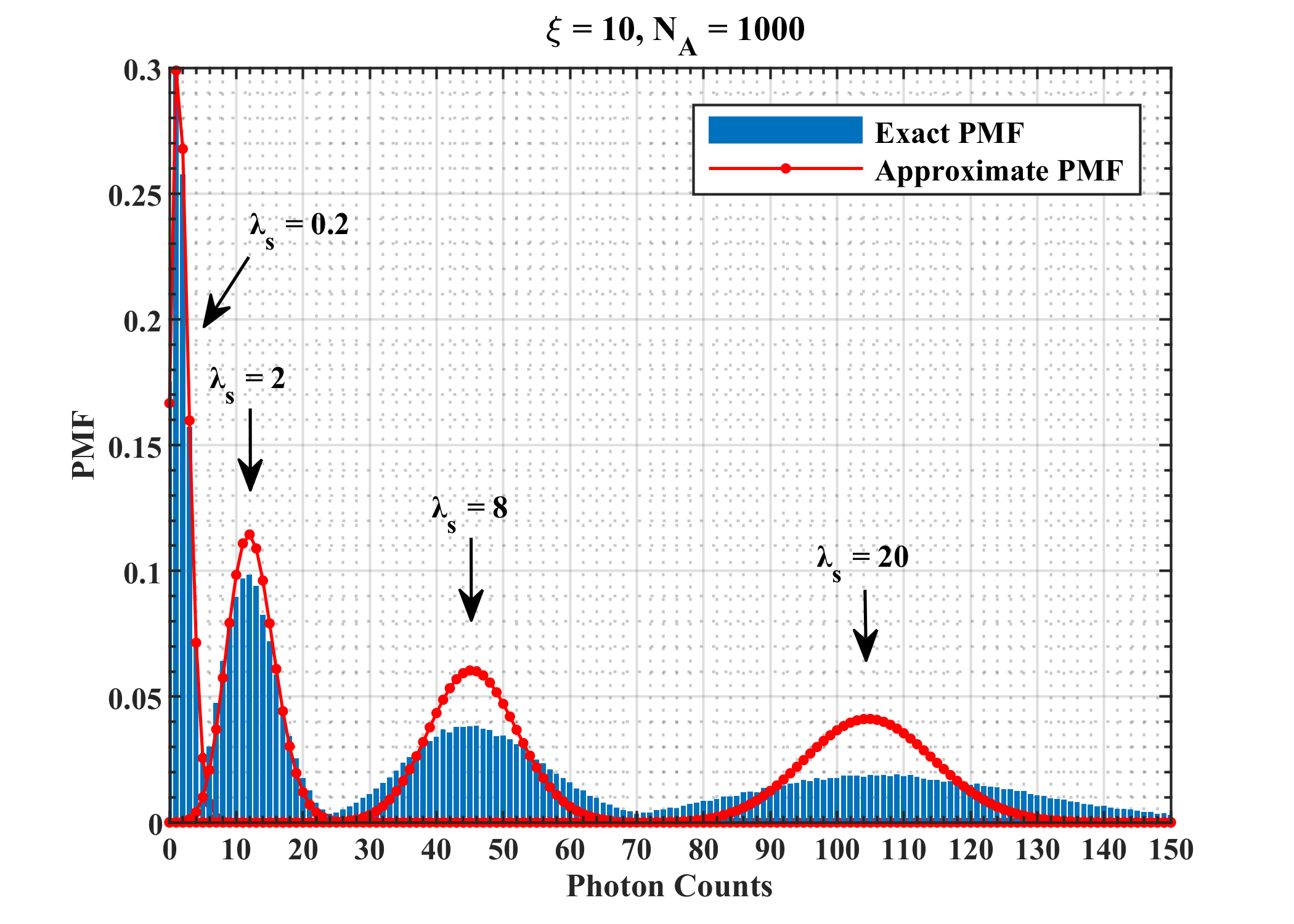}
\label{fig_5_4}}
\caption{Exact vs. approximate PMFs of detected photon counts for low-, medium-, and high-speed systems (${\lambda _{\rm{s}}} = 0.2,2,8,20 \hspace{0.2em} \rm{c/ns}$).}
\label{fig_5}
\end{figure*}

Utilizing the derived PMFs (\ref{eq15}) and (\ref{eq31}), we investigate the corresponding photon-counting statistics. To maintain the counting rate within a specific order of magnitude, the array scale is increased proportionally with the dead time ratio. Fig. \ref{fig_5} illustrates the discrepancy between the exact PMFs (obtained via Monte-Carlo simulation) and our analytical approximations under the 4-PAM scheme, whose constellation points are ${\lambda _{\rm{s}}} = 0.2,2,8,20$ counts/nanosecond ($\rm{c/ns}$). In practice, the approximate trigger probability represents a statistical estimation of the PMFs for consecutive identical symbols, leading to a mismatch between the exact and approximate PMFs. As evident in Fig. \ref{fig_5_2} and Fig. \ref{fig_5_3}, this discrepancy manifests as a lower estimation in the high-count region and a higher estimation in the low-count region.

For low- and medium-speed systems (Fig. \ref{fig_5_1} and \ref{fig_5_2}), the profiles of the approximate and exact PMFs align closely across both low- and high-count regions. In contrast, for high-speed systems (Fig. \ref{fig_5_3} and \ref{fig_5_4}), the profiles align well in the low-count region. However, a mismatch emerges in the high-count region between the approximate PMF and the Monte-Carlo simulation results. This is because, under high-order modulation with long dead time, the exact PMF of photon counts comprises a superposition of binomial distributions. Consequently, compared to the approximate PMF, the exact PMF exhibits broadening and can develop multiple peaks (as seen in Fig. \ref{fig_5_3} ). 

To derive a closed-form PMF, we simplified the process as a single binomial distribution. Although the peak structures differ substantially, the distribution tails exhibit high similarity, which is crucial for SER assessment and threshold computation. Therefore, the approximated PMF remains effective for symbol error performance analysis. These findings demonstrate that the analytical model accurately predicts the counting statistics for SPAD array-based photon-counting receivers, even under conditions of long dead time.

\section{SPAD Array-based Optical Wireless Communication Systems}
In this section, we investigate the impact of dead time on the symbol error performance of a SPAD array-based optical communication system employing PAM. Optical communication systems require the SPAD receiver to operate over consecutive symbol intervals. Within such systems, the dead time affects the photon-counting process in two primary ways.

The first effect is the introduction of counting losses within each symbol interval. The second effect stems from the potential temporal overlap of dead time between adjacent intervals. Specifically, the dead time triggered by the last detected photon in one symbol interval may extend into the next one or several subsequent intervals. This results in a temporary blockage of the SPAD at the beginning of those intervals, causing additional counting losses and distorting the recorded photon counts. We refer to this phenomenon as dead-time-induced ISI. Consequently, the SPAD's dead time significantly constrains the maximum achievable data rate of the system. Commercially available SPAD devices exhibit dead times ranging from a few to tens of ns.

In $M$-PAM, each symbol is transmitted by sending one of $M$ distinct optical intensity pulses over a symbol duration. The receiver employs direct detection, where the incoming optical signal is incident upon the SPAD array. The number of avalanche events registered by the SPAD array within each interval is processed to determine which of the $M$ pulses was received, thereby recovering the transmitted symbol.

Modeling the SPAD array as a communication channel, we adopt a discrete-time signaling scheme: the channel input is the instantaneous optical signal intensity, which remains constant over discrete intervals but can vary between them. The channel output is the photon counts recorded by the SPAD array in each interval. Let $\chi  = \left\{ {{x_1}, \ldots ,{x_m}, \ldots ,{x_M}} \right\}$ represent the constellation points of the $M$-PAM transmitter. Thus, ${\lambda _{\rm{s}}}\left( {{x_m}} \right)$ denotes the average signal photon rate corresponding to symbol "${x_m}$". Consequently, the average total carrier rate (including both photon-generated and dark carriers) per pixel for symbol "${x_m}$" is given by $\lambda \left( {{x_m}} \right) \triangleq \left[ {{{{p_{\rm{d}}}\left( {{\lambda _{\rm{s}}}\left( {{x_m}} \right) + {\lambda _{\rm{b}}}} \right)} \mathord{\left/
 {\vphantom {{{p_{\rm{d}}}\left( {{\lambda _{\rm{s}}}\left( {{x_m}} \right) + {\lambda _{\rm{b}}}} \right)} {{N_{{\rm{A}}}}}}} \right.
 \kern-\nulldelimiterspace} {{N_{{\rm{A}}}}}}} \right] + {\lambda _{\rm{d}}}$.
 
\subsection{SER in Low- and Medium-Speed Systems}
For low- and medium-speed systems, the symbol duration is sufficiently long that each SPAD pixel can potentially record more than one photon counts. In this regime, the photon-counting process is modeled by a renewal process. Consequently, the PMF of the total photon counts registered by the SPAD array (i.e., the summation across all pixels) follows a multinomial distribution (\ref{eq15}). In $M$-PAM, information symbols are conveyed by modulating the intensity of the optical signal. Therefore, the probability that exactly $k$ photons are counted by the entire SPAD array within the symbol duration ${T_{\rm{s}}}$, when symbol  ${x_m}$ is transmitted, is given by:
\begin{align}\label{eq32}
{p_{\rm{e}}^{\rm{L}}}\left( {k,{x_m}} \right) = p_{{\rm{A}}}^{\rm{L}}\left( {k,{T_{\rm{s}}},\lambda\left( {{x_m}} \right)} \right)  
\end{align}

In this case, the likelihood ratio (LLR) between symbol ${x_{m + 1}}$ and symbol "${x_m}$", given the observed photon counts $k$, is defined as:
\begin{dmath}\label{eq33}
{\Lambda ^{\rm{L}}}\left( k \right) = \frac{{{p_{\rm{e}}^{\rm{L}}}\left( {k,{x_{m + 1}}} \right)}}{{{p_{\rm{e}}^{\rm{L}}}\left( {k,{x_m}} \right)}} 
\end{dmath}

Consequently, the SER under ML detection is expressed as:
\begin{dmath}\label{eq34}
P_{\rm{e}}^{\rm{L}} = \frac{1}{M}\sum\limits_{m = 1}^{M - 1} {\left\{  { \begin{array}{*{20}{c}} \sum\limits_{\left\{ {k:{\Lambda ^{\rm{L}}}\left( k \right) \le 1} \right\}} {{p_{\rm{e}}^{\rm{L}}}\left( {k,{x_{m + 1}}} \right)} \\ + \sum\limits_{\left\{ {k:{\Lambda ^{\rm{L}}}\left( k \right) > 1} \right\}} {{p_{\rm{e}}^{\rm{L}}}\left( {k,{x_m}} \right)} \end{array}} \right\}}  
\end{dmath}

The SER, denoted ${P_{\rm{e}}}$, critically depends on the detection thresholds. The optimal thresholds, selected to minimize ${P_{\rm{e}}}$, are generally challenging to derive in closed-form. However, for the case of relatively long dead time ($\xi  < 0.1$), ISI effects become negligible. Under this long-dead-time assumption, an SPAD array comprising ${N_{{\rm{A}}}}$ pixels can be effectively modeled as a single SPAD with an equivalent, prolonged symbol duration of ${T_{\rm{s}}}{N_{{\rm{A}}}}$. Consequently, the optimal thresholds can be expressed in closed-form as derived in \cite{ref28}:
\begin{align}\label{eq35}
k_{{\rm{th}}}^{\rm{L}}\left( {{x_m}} \right) = \frac{{\left[ {\lambda \left( {{x_{m + 1}}} \right) - \lambda \left( {{x_m}} \right)} \right]\left( {{T_{\rm{s}}}{N_{{\rm{A}}}} - {\tau _{\rm{d}}}} \right)}}{{\left[ {\lambda \left( {{x_{m + 1}}} \right) - \lambda \left( {{x_m}} \right)} \right]{\tau _{\rm{d}}} + \ln \left[ {\frac{{\lambda \left( {{x_{m + 1}}} \right)}}{{\lambda \left( {{x_m}} \right)}}} \right]}}   
\end{align}

Hereinafter, we adopt threshold detection for error probability calculations. The received photon counts $k$ is compared against a series of thresholds $\left\{ {k_{{\rm{th}}}^{\rm{L}}\left( {{x_1}} \right),k_{{\rm{th}}}^{\rm{L}}\left( {{x_2}} \right), \cdots ,k_{{\rm{th}}}^{\rm{L}}\left( {{x_{M - 1}}} \right)} \right\}$ to decide the transmitted symbol. For equally likely symbols, the SER is then given by:
\begin{dmath}\label{eq36}
P_{\rm{e}}^{\rm{L}} = \frac{1}{M}\sum\limits_{m = 1}^{M - 1} {\left\{ {\begin{array}{*{20}{c}}\sum\limits_{k = 0}^{\left\lfloor {k_{{\rm{th}}}^{\rm{L}}\left( {{x_m}} \right)} \right\rfloor } {p_{\rm{e}}^{\rm{L}}}\left( {k,{x_{m + 1}}} \right) \\ + \sum\limits_{k = \left\lceil {k_{{\rm{th}}}^{\rm{L}}\left( {{x_m}} \right)} \right\rceil }^{{k_{\max }}} {{p_{\rm{e}}^{\rm{L}}}\left( {k,{x_m}} \right)} \end{array}} \right\}}   
\end{dmath}
where $\left\lfloor  \cdot  \right\rfloor$ denotes the floor function.

Although (\ref{eq34}) is derived at low-speed conditions ($
\xi  < 0.1$), it can also be applied to medium-speed systems, which is close to the optimal threshold of ML detection. The threshold detection of medium-speed is also made by (\ref{eq35}) in the context.

\subsection{SER in High-Speed Systems}
For the high-speed system, the probability that exactly $k$ photons are counted by the entire SPAD array within the symbol duration ${T_{\text{s}}}$, when symbol "${x_m}$" is transmitted, is:
\begin{align}\label{eq37}
{p_{\rm{e}}^{\rm{H}}}\left( {k,{x_m}} \right) = p_{{\rm{A}}}^{\rm{H}}\left( {k,{T_{\rm{s}}},{\lambda _{\rm{s}}}\left( {{x_m}} \right)} \right)   
\end{align}

Then, we also derive the SER under ML detection as:
\begin{align}\label{eq38}
P_{\rm{e}}^{\rm{H}} = \frac{1}{M}\sum\limits_{m = 1}^{M - 1} {\left\{ {\sum\limits_{\left\{ {k:{\Lambda ^{\rm{H}}} \le 1} \right\}} {{p_{\rm{e}}^{\rm{H}}}\left( {k,{x_{m{\rm{ + }}1}}} \right)}  + \sum\limits_{\left\{ {k:{\Lambda ^{\rm{H}}} > 1} \right\}} {{p_{\rm{e}}^{\rm{H}}}\left( {k,{x_m}} \right)} } \right\}}  
\end{align}
where the LLR is expressed:
\begin{align}\label{eq39}
{\Lambda ^{\rm{H}}}\left( k \right) = \frac{{{p_{\rm{e}}^{\rm{H}}}\left( {k,{x_{m{\rm{ + }}1}}} \right)}}{{{p_{\rm{e}}^{\rm{H}}}\left( {k,{x_m}} \right)}}  
\end{align}

To facilitate analysis, we approximate the PMF given in (\ref{eq31}) as a binomial distribution. According to ((15), \cite{ref37}), the threshold is expressed as:
\begin{dmath}\label{eq40}
k_{{\rm{th}}}^{\rm{H}}\left( {{x_m}} \right) = \frac{{{k_{\max }}\ln \left[ {\frac{{1 - {p_{{\rm{apr}}}}\left( {{x_m}} \right)}}{{1 - {p_{{\rm{apr}}}}\left( {{x_{m + 1}}} \right)}}} \right]}}{{\ln \left[ {\frac{{{p_{{\rm{apr}}}}\left( {{x_{m + 1}}} \right)\left( {1 - {p_{{\rm{apr}}}}\left( {{x_m}} \right)} \right)}}{{{p_{{\rm{apr}}}}\left( {{x_m}} \right)\left( {1 - {p_{{\rm{apr}}}}\left( {{x_{m + 1}}} \right)} \right)}}} \right]}}    
\end{dmath}
where ${p_{{\rm{apr}}}}\left( {{x_m}} \right) \triangleq {p^{\rm{H}}}\left( {1,{T_{\rm{s}}},\lambda\left( {{x_m}} \right)} \right){p_{{\rm{S}}}}\left( 1 \right)$ represents the equivalent probability that a SPAD pixel detects a photon within the symbol "${x_m}$".

Then, for equally likely symbols, the SER using threshold detection based on this approximation is expressed as:
\begin{align}\label{eq41}
P_{\rm{e}}^{\rm{H}} = \frac{1}{M}\sum\limits_{m = 1}^{M - 1} {\left\{ {\begin{array}{*{20}{c}}\sum\limits_{k = 0}^{\left\lfloor {k_{{\rm{th}}}^{\rm{H}}\left( {{x_m}} \right)} \right\rfloor } {p_{\rm{e}}^{\rm{H}}}\left( {k,{x_{m{\rm{ + }}1}}} \right) \\ + \sum\limits_{k = \left\lceil {k_{{\rm{th}}}^{\rm{H}}\left( {{x_m}} \right)} \right\rceil }^{{k_{\max }}} {{p_{\rm{e}}^{\rm{H}}}\left( {k,{x_m}} \right)}  \end{array}} \right\}}     
\end{align}

\subsection{ML Detection vs. Threshold Detection}
This section presents symbol error performance comparisons, juxtaposing analytical results with Monte-Carlo simulations. Across all figures, we compare the analytical SER against Monte-Carlo simulation results employing both ML detection and threshold detection. To derive a closed-form PMF formula, we approximate the photon counting process incorporating dead time. This approximation inevitably introduces a mismatch in the distribution tails of the PMF, leading to an SER gap between the analytical model and Monte-Carlo simulations. For low- and medium-speed systems, we introduce a critical dead time ratio ($\xi {\rm{ = 0}}{\rm{.1}}$). For high-speed systems, a long dead time ratio ($\xi {\rm{ = }}10$) is employed. Results obtained from the analytical model are labeled "\textbf{Ana.}", while those from Monte-Carlo simulations are labeled "\textbf{Sim.}". Signal demodulation using ML detection is denoted "\textbf{ML}", and demodulation using threshold detection is denoted "\textbf{TH}".

\begin{figure*}[!htb]
\centering
\hfil
\subfloat[Low-speed]{\includegraphics[width=3.3in]{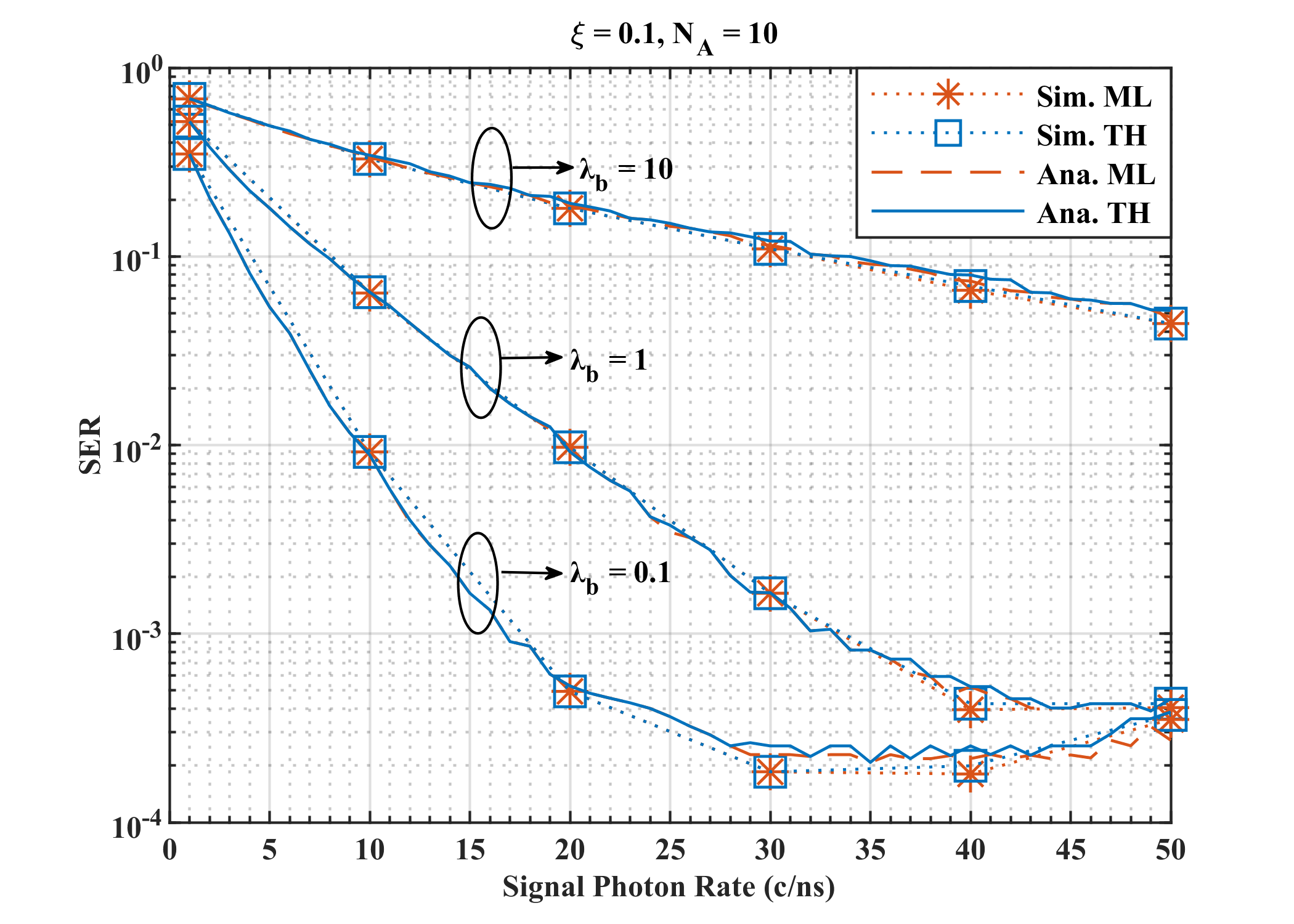}
\label{fig_6_1}}
\hfil
\subfloat[High-speed]{\includegraphics[width=3.3in]{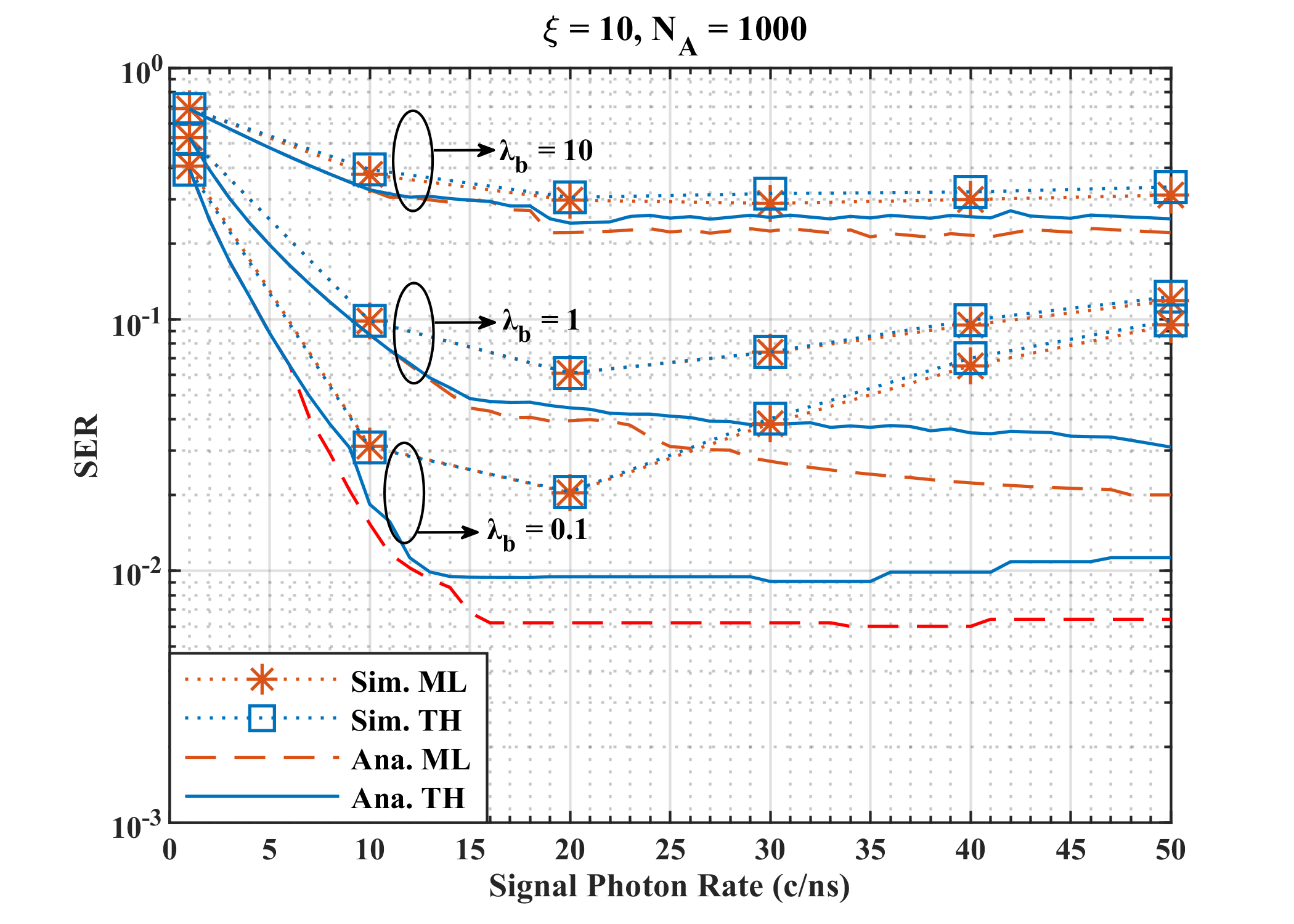}
\label{fig_6_2}}
\caption{Simulated vs. analytical SER versus received signal photon rate for SPAD array-based OWC systems.}
\label{fig_6}
\end{figure*}

To evaluate the validity of the analytical model and explore system performance under various SPAD operating conditions, signal photon rates range from $1\!\sim\!50\hspace{0.2em} \rm{c/ns}$. As shown in Fig. \ref{fig_6}, we observe remarkable agreement between the SER performance of ML detection and threshold detection across all cases. This agreement underscores that the thresholds defined by (\ref{eq35}) and (\ref{eq40}) closely approximate the optimal ML detection performance. Furthermore, the analytical SER results align well with the Monte-Carlo simulations, with discrepancies between them consistently below one order of magnitude. Therefore, these results demonstrate the validity of the analytical model for low-, medium-, and high-speed systems utilizing practical photon-counting receivers.

In Fig. \ref{fig_6}, The SER initially decreases at low photon rates but increases once the rate exceeds approximately $20\hspace{0.2em} \rm{c/ns}$. In low-photon-rate regimes, increasing the signal photon rate primarily drives SER improvement. However, beyond a certain threshold, photon counts saturation occurs, degrading the SER. High-speed systems exhibit a lower saturation level than low-speed systems. This is because their broader PMF is more sensitive to increasing signal photon rates, leading to greater overlap in the distributions. Consequently, maintaining an appropriate incident optical intensity is critical for photon-counting receivers.

\section{Numerical Results and Discussions}
This section investigates the symbol error performance of photon-counting receivers. In simulations, the dead time is set on the order of one to tens of nanoseconds, consistent with practical SPAD devices. The background photon rate is measured as ${\lambda _{\rm{b}}} = 0.1\!\sim\!10\hspace{0.2em} \rm{c/ns}$, based on prior experimental data \cite{ref38}. To mitigate the inherent signal-dependent noise in photon-counting channels, we employ an unequally spaced 4-PAM constellation designed using the square-root signaling method \cite{ref39,ref40,ref41}. Building upon the square-root signaling foundation, we modify the PAM constellation, resulting in the final signal levels $\left\{{0,0.1{\lambda_{\rm{s}}},0.4{\lambda_{\rm{s}}},{\lambda_{\rm{s}}}}\right\}$ for 4-PAM. The results are obtained by applying threshold detection to both the analytical model and Monte-Carlo simulation outputs.

\begin{figure*}[!htb]
\centering
\hfil
\subfloat[Low-speed]{\includegraphics[width=3.3in]{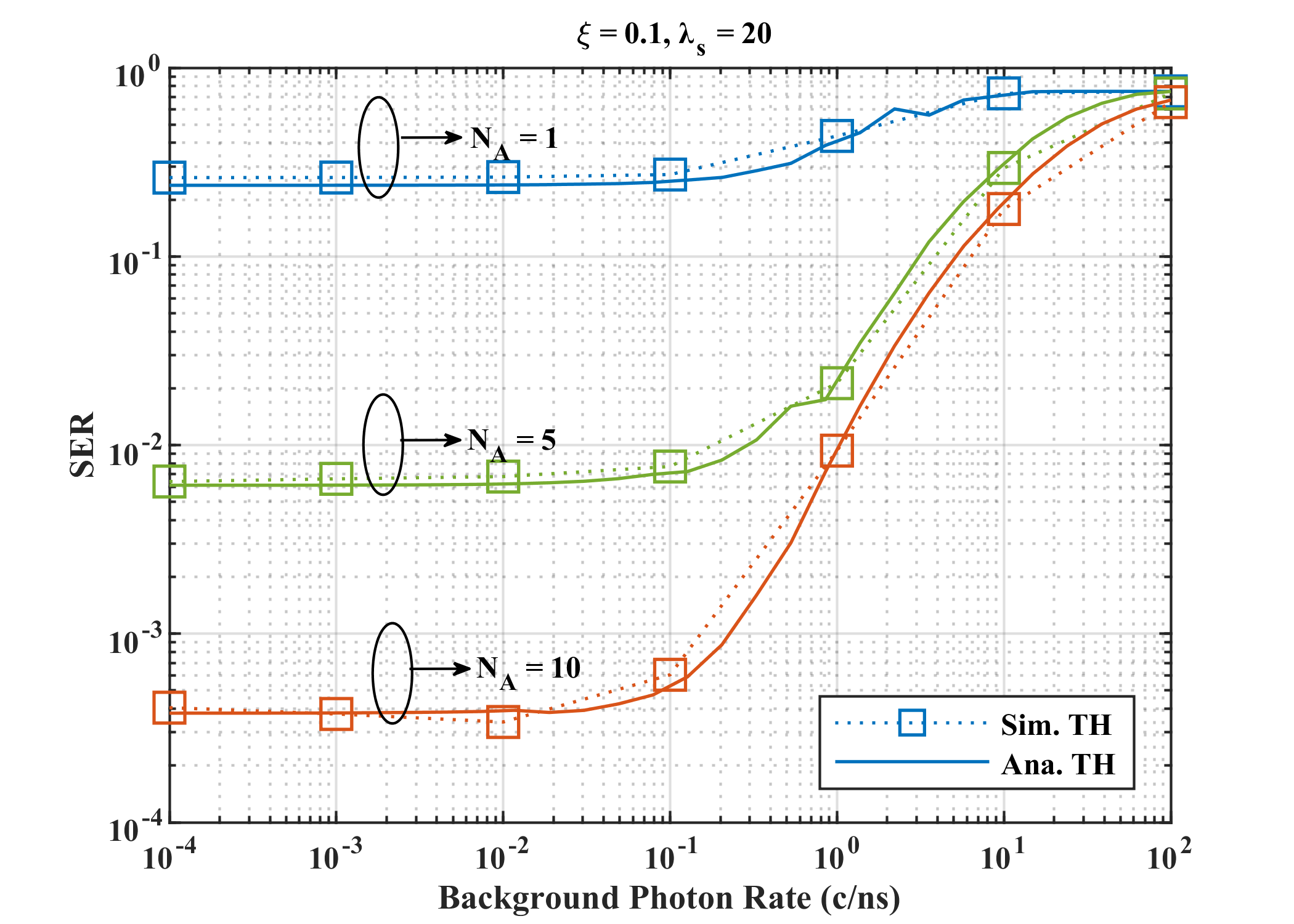}
\label{fig_7_1}}
\hfil
\subfloat[High-speed]{\includegraphics[width=3.3in]{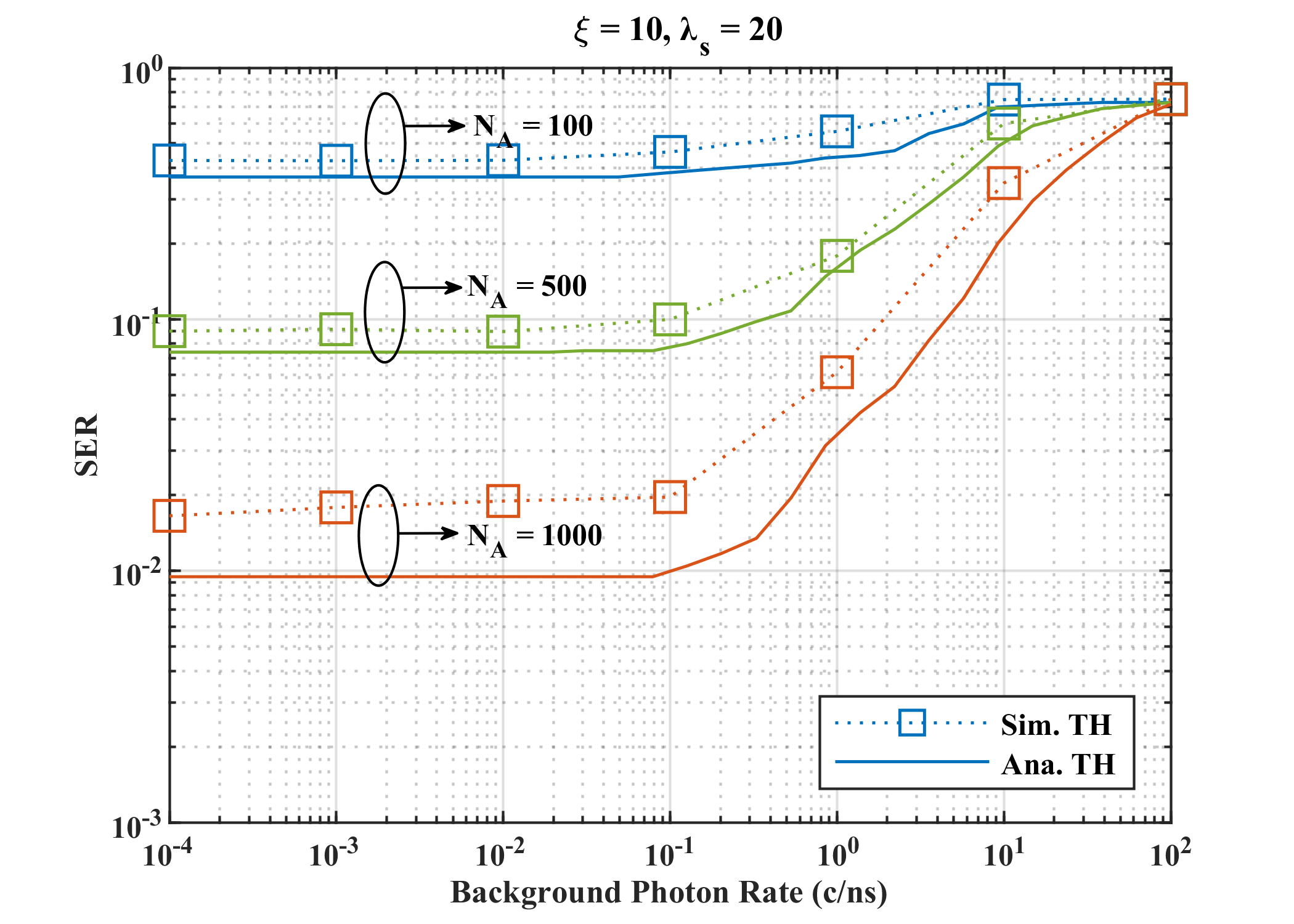}
\label{fig_7_2}}
\caption{SER versus background photon rate under different array scales for low- and high-speed systems.}
\label{fig_7}
\end{figure*}

Fig. \ref{fig_7} investigates the impact of background radiation across the range of $10^{-4} \!\sim\! 10^2\hspace{0.2em} \rm{c/ns}$, encompassing most practical communication scenarios. The performance of practical photon-counting receivers is highly dependent on background radiation. Strong background radiation degrades error performance as the SPAD saturates with false photon counts generated by background photons. As shown in Fig. \ref{fig_7}, SER degrades rapidly across all array sizes when the background photon rate exceeds $0.1\hspace{0.2em} \rm{c/ns}$. This upper limit defines the tolerated background radiation level. Clearly, maintaining the background radiation below this tolerated level is essential for optimal system performance.

\begin{figure*}[!htb]
\centering
\hfil
\subfloat[Low-speed]{\includegraphics[width=3.3in]{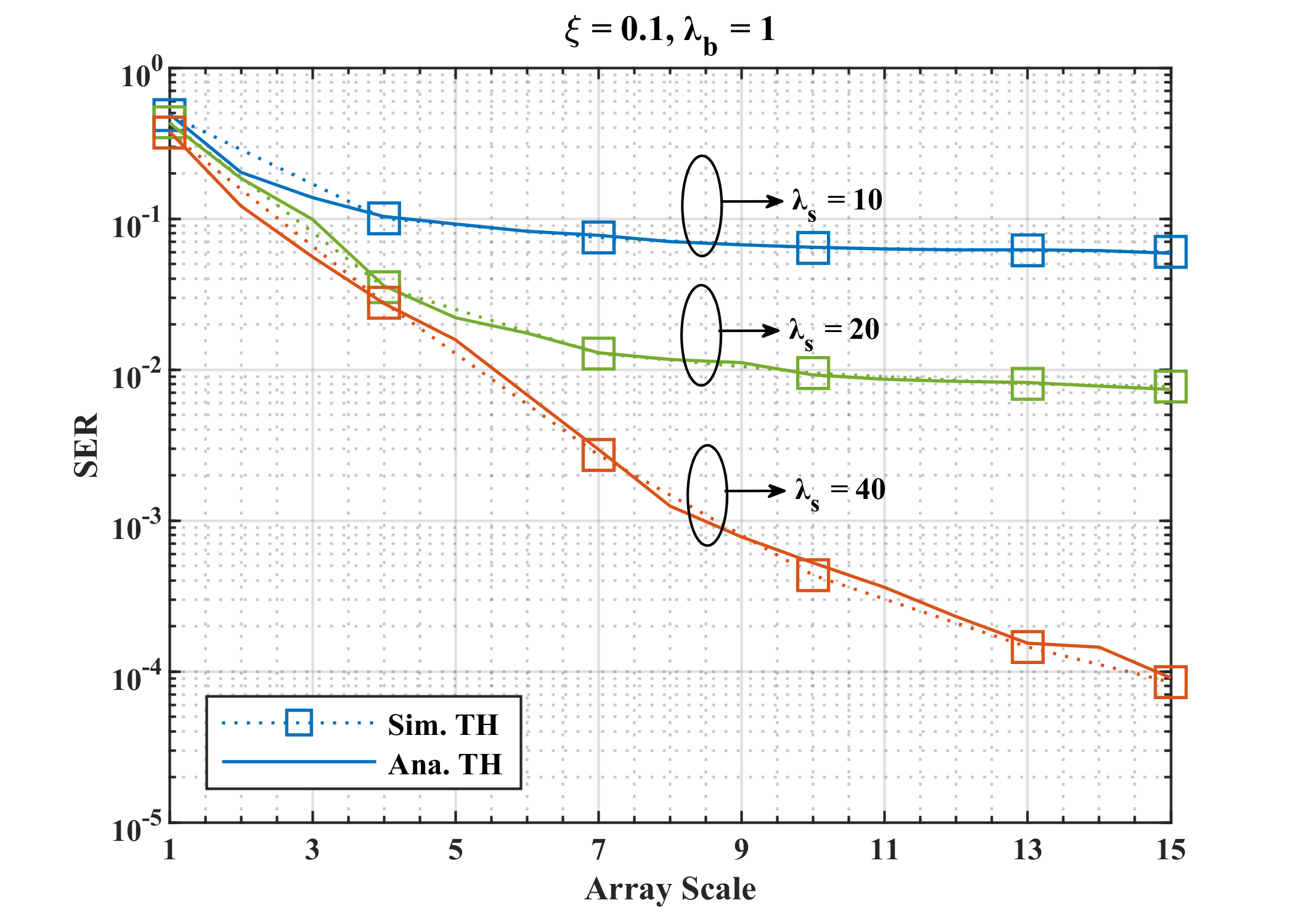}
\label{fig_8_1}}
\hfil
\subfloat[High-speed]{\includegraphics[width=3.3in]{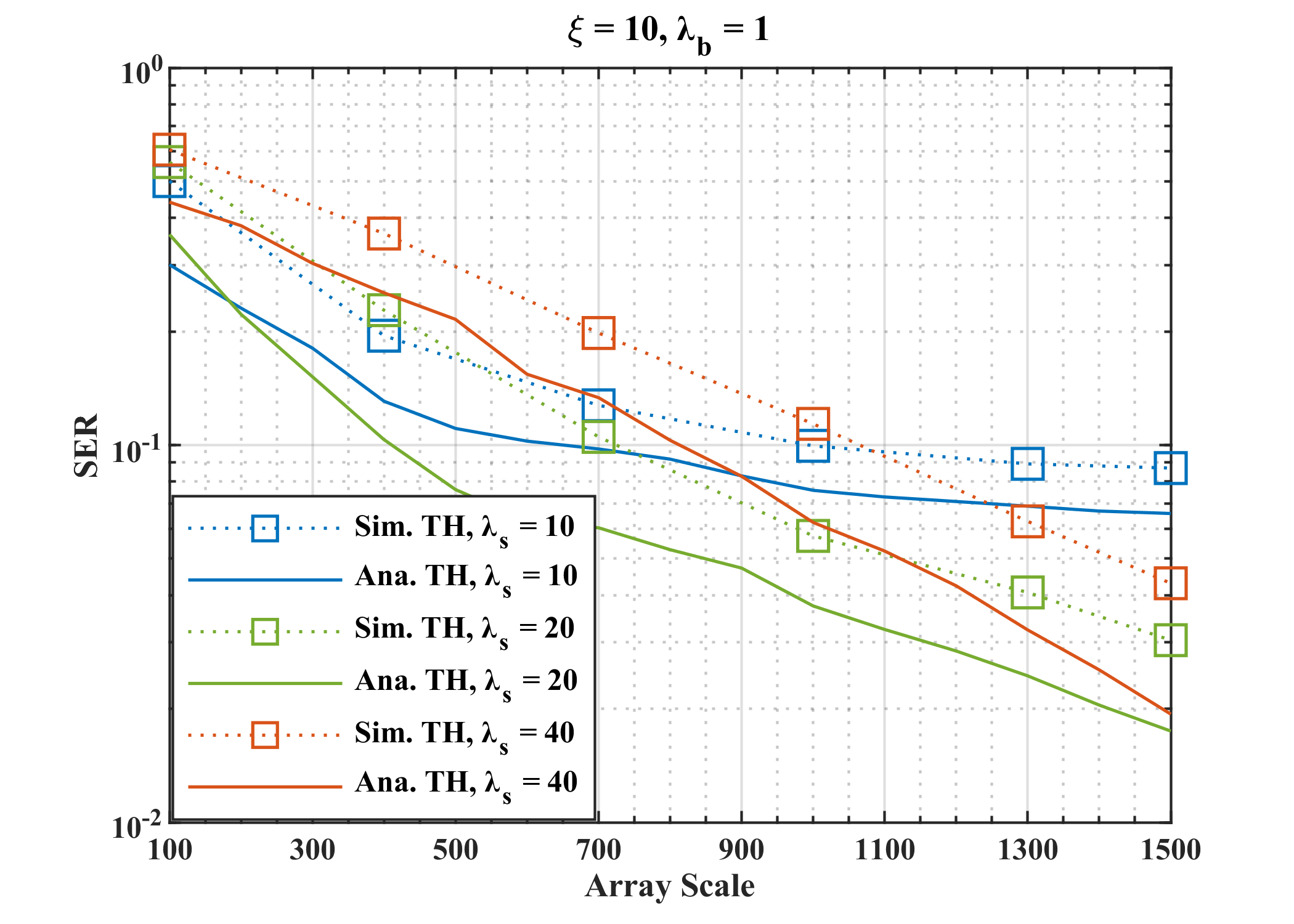}
\label{fig_8_2}}
\caption{SER versus array scale at different signal photon rates for low- and high-speed systems.}
\label{fig_8}
\end{figure*}

Fig. \ref{fig_8} presents SER results as a function of array size. SER is observed to decrease exponentially with increasing array size. Larger arrays enhance the maximum count rate, thereby distributing the PMF more broadly over the possible photon counts. Increasing the array size is consistently an effective method for improving error performance in SPAD array-based receivers. Therefore, the array size dominates the SER floor when both the optical signal and background radiation intensities are maintained at appropriate levels. This effect is particularly crucial for high-speed systems. In such systems, where each pixel is typically limited to outputting only one photon counts per symbol period, the discrete nature of the counts necessitates a larger array size to sufficiently separate the PMFs corresponding to different symbols and thus improve error performance. Thus, maintaining a sufficient array size significantly enhances the achievable count rate and optimizes error performance.

\begin{figure*}[!htb]
\centering
\hfil
\subfloat[Constant symbol duration]{\includegraphics[width=3.3in]{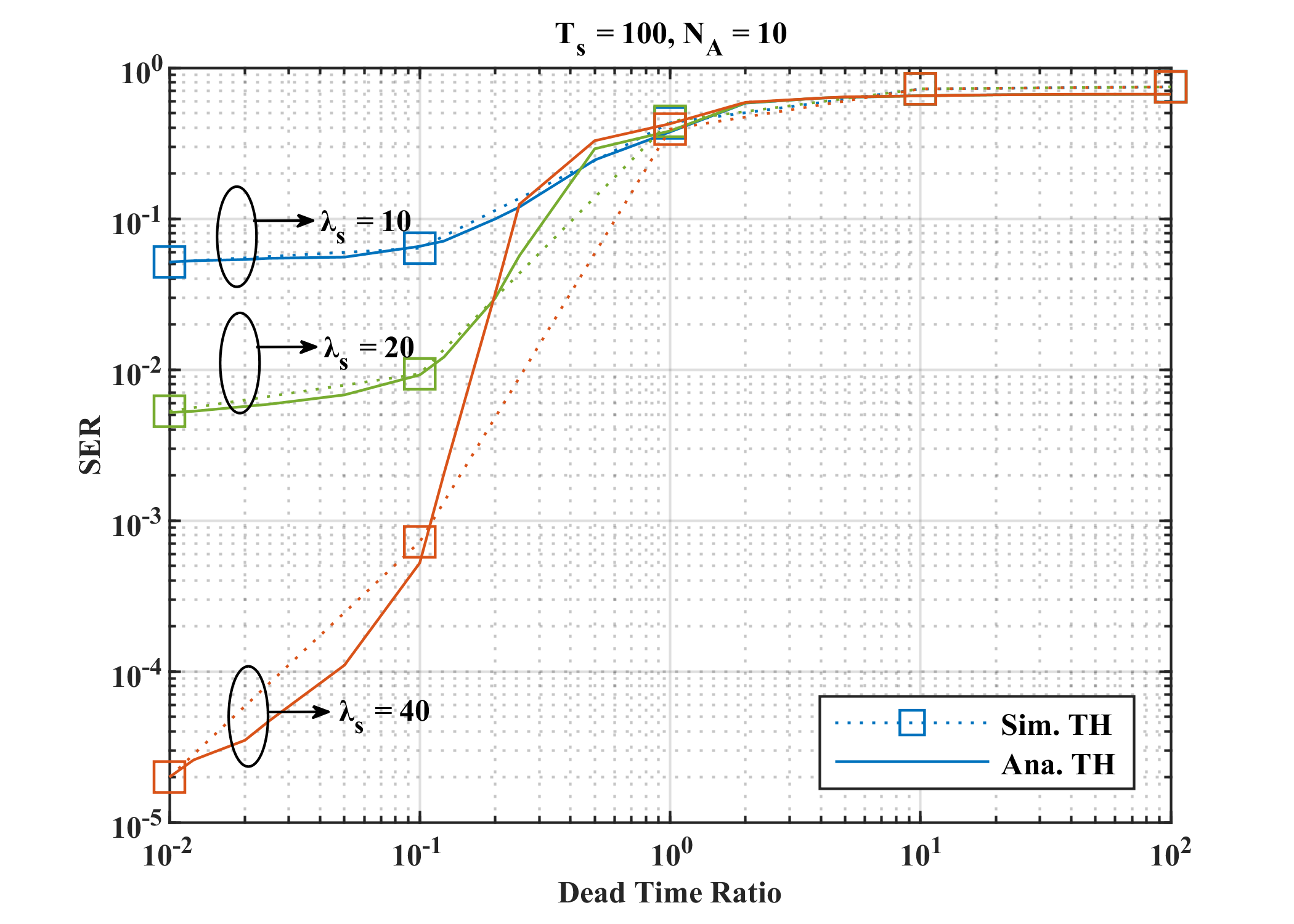}
\label{fig_9_1}}
\hfil
\subfloat[Constant dead time]{\includegraphics[width=3.3in]{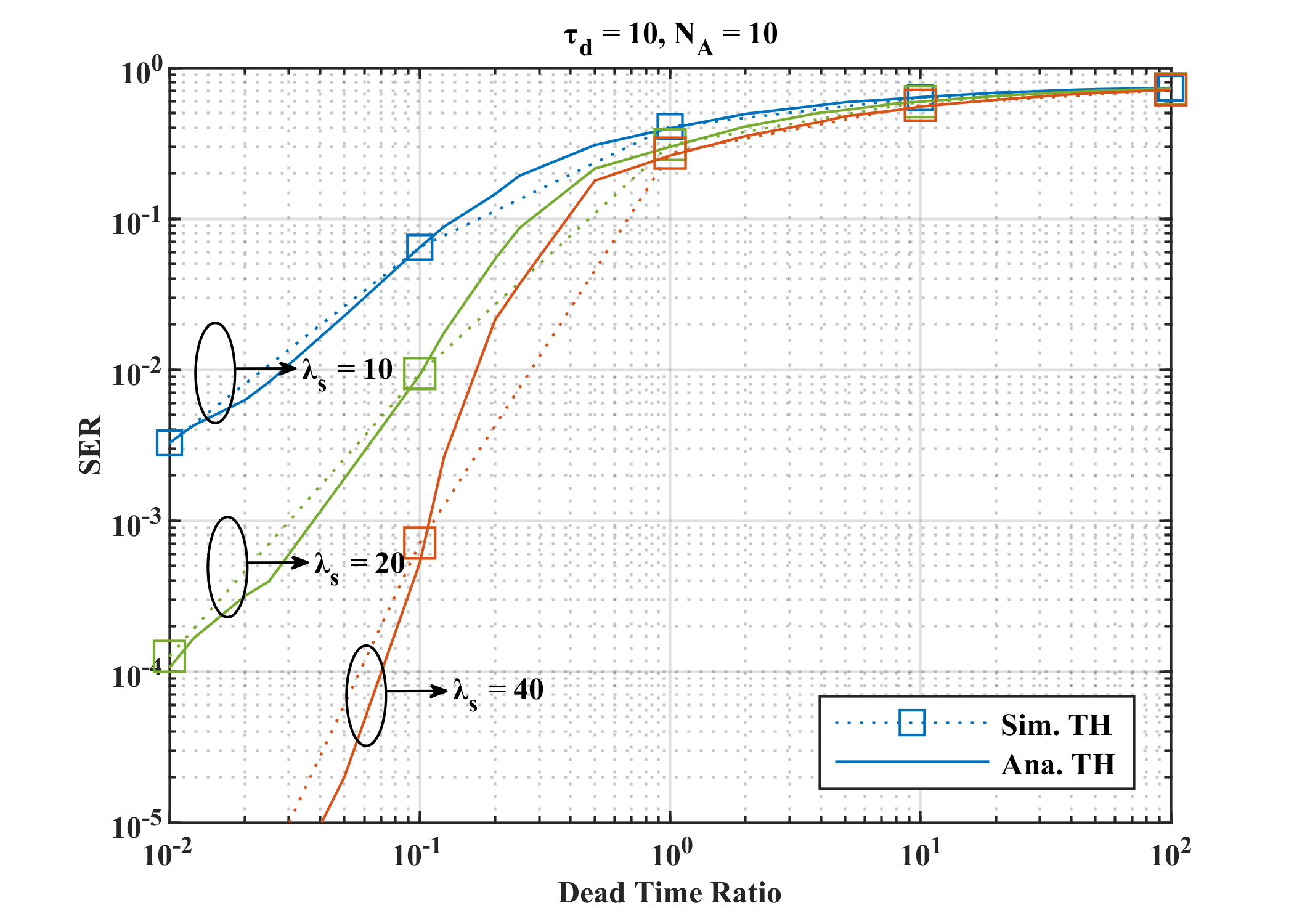}
\label{fig_9_2}}
\caption{SER versus dead time ratio at different signal photon rates for constant symbol duration and dead time.}
\label{fig_9}
\end{figure*}

Fig. \ref{fig_9} depicts the SER as a function of increasing dead time ratio, which can be achieved either by extending the dead time or by shortening the symbol duration. The results show that the SER increases exponentially with rising dead time ratio (corresponding to a reduced count rate). When the dead time ratio is increased by prolonging the dead time, the SER rises across all optical signal regimes. In contrast, when the dead time ratio is increased by shortening the symbol duration while keeping the dead time constant, the error performance also exhibits degradation. At low dead time ratios ($\xi  < 1$), the SER observed under the constant dead time condition is lower than that under the constant symbol duration condition. In this scenario, as the dead time ratio decreases, the symbol duration increases. This improvement in SER is attributed to the enhanced optical signal power. Consequently, the achievable dead time ratio is constrained by both the optical signal intensity and the target SER requirement.

In summary, the fundamental error performance characteristics of low- and high-speed systems share similarities. However, as discussed, high-speed systems require a large SPAD array to maintain an adequate count rate. Due to the broader PMF associated with larger arrays, the saturation optical intensity for high-speed systems is lower. Furthermore, the tolerated background radiation level is nearly identical for both system types. Based on this analysis, the key factors for minimizing SER are: (1) maintaining background radiation below its tolerated level; (2) ensuring an adequate count rate (primarily governed by array size); (3) operating the signal intensity within an appropriate range to avoid saturation. Finally, the available dead time ratio is determined by the optical signal intensity and the target SER requirement.

\section{Conclusion}
In conclusion, we have developed a comprehensive analytical framework for modeling the statistical behavior of practical photon-counting receivers incorporating dead time. Our analysis quantifies the impact of dead time-induced blocking loss and ISI on photon-counting statistics, leading to a closed-form approximation for the photon counts probability distribution. Furthermore, we derived closed-form threshold expressions based on the ML criterion. The proposed statistical model accurately captures the detected photon counts for high-order PAM signals under dead time constraints in both low- and high-speed systems. Utilizing this analytical model enables efficient design and optimization of photon-counting receiver parameters. Overall, this study provides a reliable analytical model for evaluating the performance of photon-limited OWC systems employing practical photon-counting receivers.

\section*{Appendix A \\PMF of the Detected Photons without Dead Time}
For weak optical signals, the arrival of incident photons follows a Poisson process. Therefore, the PMF of the number of incident signal and background photons within a time interval $\left( {0,t} \right)$ is presented as:
\begin{align}\label{eq42}
P\left\{ {{N_{\rm{s}}}\left( t \right) = k} \right\} = \frac{{{{\left[ {\left( {{\lambda _{\rm{s}}} + {\lambda _{\rm{b}}}} \right)t} \right]}^k}}}{{k!}}{{\rm{e}}^{ - \left( {{\lambda _{\rm{s}}} + {\lambda _{\rm{b}}}} \right)t}}  
\end{align}

Assuming an ideal SPAD operating without dead time, each incident photon is detected independently with a probability equal to the PDE, denoted ${p_{\rm{d}}}$. Consequently, the detection process can be modeled as binomial selection applied to the incident Poisson stream. The PMF of the number of detected photons (arising solely from photon-generated carriers, before considering dark counts), conditioned on $i$ incident photons, is binomial $\mathbb{B} \left( {i,{p_{\rm{d}}}} \right)$. The PMF of detected photons without dead time can be derived as:
\begin{dmath}\label{eq43}
P\left\{ {{N_{\det }}\left( t \right) = k} \right\} = \sum\limits_{i = 1}^{ + \infty } {P\left\{ {\mathbb{B} = k,{N_{\rm{s}}}\left( t \right) = i} \right\}} \\
 = \sum\limits_{i = n}^{ + \infty } {P\left\{ {\mathbb{B} = k\left| {{N_{\rm{s}}}\left( t \right) = i} \right.} \right\}P\left\{ {{N_{\rm{s}}}\left( t \right) = j} \right\}} \\
 = \sum\limits_{i = k}^{ + \infty } \binom{i}{k}{p_{\rm{d}}}^k{{\left( {1 - {p_{\rm{d}}}} \right)}^{i - k}}\frac{{{{\left[ {\left( {{\lambda _{\rm{s}}} + {\lambda _{\rm{b}}}} \right)t} \right]}^i}}}{{i!}}{e^{ - {\lambda _{\rm{p}}}t}} \\
 = \frac{{{{\left[ {{p_{\rm{d}}}\left( {{\lambda _{\rm{s}}} + {\lambda _{\rm{b}}}} \right)t} \right]}^k}}}{{k!}}{{\rm{e}}^{ - \left( {{\lambda _{\rm{s}}} + {\lambda _{\rm{b}}}} \right)t}}\underbrace {\sum\limits_{i = 0}^{ + \infty } {\frac{{{{\left[ {\left( {1 - {p_{\rm{d}}}} \right)\left( {{\lambda _{\rm{s}}} + {\lambda _{\rm{b}}}} \right)t} \right]}^i}}}{{i!}}} }_{{\rm{ = }}{{\rm{e}}^{ - \left( {1 - {p_{\rm{d}}}} \right)\left( {{\lambda _{\rm{s}}} + {\lambda _{\rm{b}}}} \right)t}}}\\
 = \frac{{{{\left[ {{p_{\rm{d}}}\left( {{\lambda _{\rm{s}}} + {\lambda _{\rm{b}}}} \right)t} \right]}^k}}}{{k!}}{{\rm{e}}^{ - \left( {{\lambda _{\rm{s}}} + {\lambda _{\rm{b}}}} \right)t}}{{\rm{e}}^{ - \left( {1 - {p_{\rm{d}}}} \right)\left( {{\lambda _{\rm{s}}} + {\lambda _{\rm{b}}}} \right)t}}\\
 = \frac{{{{\left[ {{p_{\rm{d}}}\left( {{\lambda _{\rm{s}}} + {\lambda _{\rm{b}}}} \right)t} \right]}^k}}}{{k!}}{{\rm{e}}^{ - {p_{\rm{d}}}\left( {{\lambda _{\rm{s}}} + {\lambda _{\rm{b}}}} \right)t}}    
\end{dmath}

Marginalizing over the incident photon distribution leads to the key result: the number of detected photons also follows a Poisson distribution, but with a reduced mean rate of ${p_{\rm{d}}}\left( {{\lambda _{\rm{s}}} + {\lambda _{\rm{b}}}} \right)$. This reduction represents photons lost during the imperfect photoelectric conversion process. Furthermore, the generation of dark carriers is also well-modeled by a Poisson arrival process with a constant average rate ${N_{\rm{d}}}\left( t \right)$. The PMF of dark carriers without dead time is:
\begin{align}\label{eq44}
P\left\{ {{N_{\rm{d}}}\left( t \right) = k} \right\} = \frac{{{{\left[ {{\lambda _{\rm{d}}}t} \right]}^k}}}{{k!}}{{\rm{e}}^{ - {\lambda _{\rm{d}}}t}}  
\end{align}

We define $N\left( t \right) \triangleq {N_{\det }}\left( t \right) + {N_{\rm{d}}}\left( t \right)$ as the total number of photon counts registered by the SPAD in the absence of dead time. This count is the sum of the counts from detected photons and dark carriers. Exploiting the additive property of independent Poisson random variables, the overall PMF of $k$ photon counts within interval $\left( {0,t} \right)$ is:
\begin{align}\label{eq45}
P\left\{ {N\left( t \right) = k} \right\} = \frac{{{{\left( {\lambda t} \right)}^k}}}{{k!}}{{\rm{e}}^{ - \lambda t}}   
\end{align}
where $\lambda  \triangleq {p_{\rm{d}}}\left( {{\lambda _{\rm{s}}} + {\lambda _{\rm{b}}}} \right) + {\lambda _{\rm{d}}}$. This completes the proof.

\section*{Appendix B \\PDF of the Last photon counts Arrival Time}
Firstly, we derive the probability that the last detected photon counts occurs before a specific time instant $t,\left( {0 \le t \le {T_{\rm{s}}}} \right)$ within the symbol duration:
\begin{dmath}\label{eq46}
P\left\{ {{T_{\rm{L}}} \le t} \right\} = \sum\limits_{k = 1}^{\left\lceil {{{{T_{\rm{s}}}} \mathord{\left/
 {\vphantom {{{T_{\rm{s}}}} {{\tau _{\rm{d}}}}}} \right.
 \kern-\nulldelimiterspace} {{\tau _{\rm{d}}}}}} \right\rceil } {P\left\{ {{T_{\rm{L}}} \le t\left| {N\left( {{T_{\rm{s}}}} \right) = k} \right.} \right\}p\left( {k,{T_{\rm{s}}}} \right)} \\
 = \sum\limits_{k = 1}^{\left\lceil {{{{T_{\rm{s}}}} \mathord{\left/
 {\vphantom {{{T_{\rm{s}}}} {{\tau _{\rm{d}}}}}} \right.
 \kern-\nulldelimiterspace} {{\tau _{\rm{d}}}}}} \right\rceil } {\frac{{P\left\{ {{T_{\rm{L}}} \le t,N\left( {{T_{\rm{s}}}} \right) = k} \right\}}}{{P\left( {k,{T_{\rm{s}}}} \right)}}p\left( {k,{T_{\rm{s}}}} \right)} \\
 = \sum\limits_{k = 1}^{\left\lceil {{{{T_{\rm{s}}}} \mathord{\left/
 {\vphantom {{{T_{\rm{s}}}} {{\tau _{\rm{d}}}}}} \right.
 \kern-\nulldelimiterspace} {{\tau _{\rm{d}}}}}} \right\rceil } {P\left\{ {{T_{\rm{L}}} \le t,N\left( {{T_{\rm{s}}}} \right) = k} \right\}} \\
 \approx \sum\limits_{k = 1}^{\left\lceil {{{{T_{\rm{s}}}} \mathord{\left/
 {\vphantom {{{T_{\rm{s}}}} {{\tau _{\rm{d}}}}}} \right.
 \kern-\nulldelimiterspace} {{\tau _{\rm{d}}}}}} \right\rceil } {p\left( {k,t} \right)p\left( {0,{T_{\rm{s}}} - t} \right)} \\
 = \left( {1 - p\left( {0,t} \right)} \right)p\left( {0,{T_{\rm{s}}} - t} \right)
\end{dmath}
where $p\left( {0,t} \right)$ refers to the probability of SPAD detecting no photons in time interval $\left( {0,t} \right)$. Based on (\ref{eq44}), it can be expressed as:
\begin{align}\label{eq47}
p\left( {0,t} \right) = P\left\{ {N\left( t \right) = 0} \right\} = {{\rm{e}}^{ - \lambda t}}  
\end{align}

Then,
\begin{align}\label{eq48}
P\left\{ {{T_{\rm{L}}} \le t} \right\} = \left( {1 - {{\rm{e}}^{ - \lambda t}}} \right){{\rm{e}}^{ - \lambda \left( {{T_{\rm{s}}} - t} \right)}}  
\end{align}

Based on (\ref{eq48}), the PDF of the last photon counts arrival time  can be calculated:
\begin{align}\label{eq49}
{f_{{T_{\rm{L}}}}}\left( t \right) = \frac{{dP\left\{ {{T_{\rm{L}}} \le t} \right\}}}{{dt}} = \lambda {{\rm{e}}^{ - \lambda \left( {{T_{\rm{s}}} - t} \right)}},0 \le t \le {T_{\rm{s}}}    
\end{align}

This completes the proof.

\section*{Appendix C \\PMF for Low- and Medium-Speed Systems ($\xi  < 1$)}
Firstly, applying the Laplace transform with respect to the time variable $t$ to (\ref{eq5}), (\ref{eq7}) and (\ref{eq8}), we obtain transformed expressions:
\begin{align}
{F_{{T_{{\rm{ISI}}}}}}\left( s \right) &= \frac{\lambda }{{s - \lambda }}\left( {{{\rm{e}}^{ - \lambda {\tau _{\rm{d}}}}} - {{\rm{e}}^{ - s{\tau _{\rm{d}}}}}} \right) + {{\rm{e}}^{ - \lambda {\tau _{\rm{d}}}}} \label{eq50} \\ 
{F_1}\left( s \right) &= \frac{\lambda }{{s + \lambda }} \label{eq51} \\
{F_2}\left( s \right) &= \frac{\lambda }{{s + \lambda }}{{\rm{e}}^{ - s{\tau _{\rm{d}}}}} \label{eq52}
\end{align}

Subsequently, utilizing the result from (\ref{eq13}), the Laplace transform of $p_{{\rm{ISI}}}^{\rm{L}}\left( {k,t} \right)$ with respect to $t$ is derived as (\ref{eq53}).
\begin{figure*}
\begin{dmath}\label{eq53}
P_{{\rm{ISI}}}^{\rm{L}}\left( {k,s} \right) = {\cal L}\left[ {p_{{\rm{ISI}}}^{\rm{L}}\left( {k,t} \right)} \right] = {\cal L}\left[ {{f_{{T_{{\rm{ISI}}}}}}\left( t \right) \otimes \int_0^t {\left[ {f_{{\rm{NISI}}}^{\rm{L}}\left( {k,t'} \right) - f_{{\rm{NISI}}}^{\rm{L}}\left( {k + 1,t'} \right)} \right]dt'} } \right] 
 = {\cal L}\left[ {{f_{{T_{{\rm{ISI}}}}}}\left( t \right)} \right]{\cal L}\left[ {\int_0^t {\left[ {f_{{\rm{NISI}}}^{\rm{L}}\left( {k,t'} \right) - f_{{\rm{NISI}}}^{\rm{L}}\left( {k + 1,t'} \right)} \right]dt'} } \right]\\
 = {\cal L}\left[ {{f_{{T_{{\rm{ISI}}}}}}\left( t \right)} \right]\frac{1}{s}\left\{ {{\cal L}\left[ {\int_0^t {f_{\rm{NISI}}^L\left( {k,t'} \right)} } \right] - {\cal L}\left[ {\int_0^t {f_{\rm{NISI}}^L\left( {k + 1,t'} \right)dt'} } \right]} \right\}\\
 = \left[ {{f_{{T_{{\rm{ISI}}}}}}\left( t \right)} \right]\frac{1}{s}\left\{ {{\cal L}\left[ {{f_1}\left( t \right) \otimes \underbrace {{f_2}\left( t \right) \otimes  \cdots  \otimes {f_2}\left( t \right)}_{k - 1}} \right] - {\cal L}\left[ {{f_1}\left( t \right) \otimes \underbrace {{f_2}\left( t \right) \otimes  \cdots  \otimes {f_2}\left( t \right)}_k} \right]} \right\}\\
 = \left[ {{f_{{T_{{\rm{ISI}}}}}}\left( t \right)} \right]\frac{1}{s}\left\{ {{\cal L}\left[ {{f_1}\left( t \right)} \right]\underbrace {{\cal L}\left[ {{f_2}\left( t \right)} \right] \cdots {\cal L}\left[ {{f_2}\left( t \right)} \right]}_{k - 1} - {\cal L}\left[ {{f_1}\left( t \right)} \right]\underbrace {{\cal L}\left[ {{f_2}\left( t \right)} \right] \cdots {\cal L}\left[ {{f_2}\left( t \right)} \right]}_k} \right\}\\
 = {F_{{T_{{\rm{ISI}}}}}}\left( s \right)\frac{1}{s}\left\{ {{F_1}\left( s \right){F_2}{{\left( s \right)}^{k - 1}} - {F_1}\left( s \right){F_2}{{\left( s \right)}^k}} \right\}    
\end{dmath}
\end{figure*}
Substituting the transformed expressions from (\ref{eq50}) (\ref{eq51}), and (\ref{eq52}) into (\ref{eq53}) yields (\ref{eq54}).
\begin{figure*}
\begin{dmath}\label{eq54}
P_{{\rm{ISI}}}^{\rm{L}}\left( {k,s} \right) = \left[ {\frac{\lambda }{{s - \lambda }}\left( {{{\rm{e}}^{ - \lambda {\tau _{\rm{d}}}}} - {{\rm{e}}^{ - s{\tau _{\rm{d}}}}}} \right) + {{\rm{e}}^{ - \lambda {\tau _{\rm{d}}}}}} \right]\frac{1}{s}\left\{ {\frac{\lambda }{{s + \lambda }}{{\left( {\frac{\lambda }{{s + \lambda }}{{\rm{e}}^{ - s{\tau _{\rm{d}}}}}} \right)}^{k - 1}} - \frac{\lambda }{{s + \lambda }}{{\left( {\frac{\lambda }{{s + \lambda }}{{\rm{e}}^{ - s{\tau _{\rm{d}}}}}} \right)}^k}} \right\}
 = \frac{\lambda }{{s - \lambda }}\left( {{{\rm{e}}^{ - \lambda {\tau _{\rm{d}}}}} - {{\rm{e}}^{ - s{\tau _{\rm{d}}}}}} \right)\left[ {\frac{{{\lambda ^k}}}{{s{{\left( {s + \lambda } \right)}^k}}}{{\rm{e}}^{ - s\left( {k - 1} \right){\tau _{\rm{d}}}}} - \frac{{{\lambda ^{k + 1}}}}{{s{{\left( {s + \lambda } \right)}^{k + 1}}}}{{\rm{e}}^{ - sk{\tau _{\rm{d}}}}}} \right] + {{\rm{e}}^{ - \lambda {\tau _{\rm{d}}}}}\left\{ {\frac{{{\lambda ^k}}}{{s{{\left( {s + \lambda } \right)}^k}}}{{\rm{e}}^{ - s\left( {k - 1} \right){\tau _{\rm{d}}}}} - \frac{{{\lambda ^{k + 1}}}}{{s{{\left( {s + \lambda } \right)}^{k + 1}}}}{{\rm{e}}^{ - sk{\tau _{\rm{d}}}}}} \right\}
 = {{\rm{e}}^{ - \lambda {\tau _{\rm{d}}}}}\left\langle {\overbrace {\frac{{{\lambda ^{k + 1}}}}{{s\left( {s - \lambda } \right){{\left( {s + \lambda } \right)}^k}}}{{\rm{e}}^{ - s\left( {k - 1} \right){\tau _{\rm{d}}}}}}^{{A_1}} - \overbrace {\frac{{{\lambda ^{k + 2}}}}{{s\left( {s - \lambda } \right){{\left( {s + \lambda } \right)}^{k + 1}}}}{{\rm{e}}^{ - sk{\tau _{\rm{d}}}}}}^{{A_2}}} \right\rangle  - \left\langle {\underbrace {\frac{{{\lambda ^{k + 1}}}}{{s\left( {s - \lambda } \right){{\left( {s + \lambda } \right)}^k}}}{{\rm{e}}^{ - sk{\tau _{\rm{d}}}}}}_{{A_3}} - \underbrace {\frac{{{\lambda ^{k + 2}}}}{{s\left( {s - \lambda } \right){{\left( {s + \lambda } \right)}^{k + 1}}}}{{\rm{e}}^{ - s\left( {k + 1} \right){\tau _{\rm{d}}}}}}_{{A_4}}} \right\rangle 
 + \underbrace {{{\rm{e}}^{ - \lambda {\tau _{\rm{d}}}}}\left\{ {\frac{{{\lambda ^k}}}{{s{{\left( {s + \lambda } \right)}^k}}}{{\rm{e}}^{ - s\left( {k - 1} \right){\tau _{\rm{d}}}}} - \frac{{{\lambda ^{k + 1}}}}{{s{{\left( {s + \lambda } \right)}^{k + 1}}}}{{\rm{e}}^{ - sk{\tau _{\rm{d}}}}}} \right\}}_{B}   
\end{dmath}    
\end{figure*}

Focusing on term $B$, which represents the Laplace transform of ${{\rm{e}}^{ - \lambda {\tau _{\rm{d}}}}}p_{{\rm{NISI}}}^{\rm{L}}\left( {k,t} \right)$, its inverse Laplace transform for the case $k < \left\lceil {{{{T_{\rm{s}}}} \mathord{\left/ {\vphantom {{{T_{\rm{s}}}} {{\tau _{\rm{d}}}}}} \right.\kern-\nulldelimiterspace} {{\tau _{\rm{d}}}}}} \right\rceil $ is:
\begin{dmath}\label{eq55}
{{\cal L}^{ - 1}}\left[ B \right] = {{\rm{e}}^{ - \lambda {\tau _{\rm{d}}}}}p_{{\rm{NISI}}}^{\rm{L}}\left( {k,t} \right)\\
 = {{\rm{e}}^{ - \lambda {\tau _{\rm{d}}}}}\left\langle {\sum\limits_{i = 0}^k {\psi \left( {i,\lambda \left[ {t - k{\tau _{\rm{d}}}} \right]} \right)}  - \sum\limits_{i = 0}^{k - 1} {\psi \left( {i,\lambda \left[ {t - \left( {k - 1} \right){\tau _{\rm{d}}}} \right]} \right)} } \right\rangle   
\end{dmath}

Crucially, we derive the following important inverse Laplace transform (\ref{eq56}).
\begin{figure*}
\begin{dmath}\label{eq56}
{{\cal L}^{ - 1}}\left[ {\frac{1}{{s\left( {s - \lambda } \right){{\left( {s + \lambda } \right)}^k}}}} \right] 
 = {{\cal L}^{ - 1}}\left[ { - \frac{1}{{{\lambda ^{k + 1}}s}} + \frac{1}{{{2^k}{\lambda ^{k + 1}}\left( {s - \lambda } \right)}} + \sum\limits_{i = 0}^{k - 1} {\left( {1 - \frac{1}{{{2^{i + 1}}}}} \right)\frac{1}{{{\lambda ^{i + 2}}{{\left( {s + \lambda } \right)}^{k - i}}}}} } \right]
 =  - \frac{1}{{{\lambda ^{k + 1}}}}\underbrace {{{\cal L}^{ - 1}}\left[ {\frac{1}{s}} \right]}_{ = u\left( t \right)} + \frac{1}{{{2^k}{\lambda ^{k + 1}}}}\underbrace {{{\cal L}^{ - 1}}\left[ {\frac{1}{{\left( {s - \lambda } \right)}}} \right]}_{ = {{\rm{e}}^{\lambda t}}}
 + \sum\limits_{i = 0}^{k - 1} {\left( {1 - \frac{1}{{{2^{i + 1}}}}} \right)\frac{1}{{{\lambda ^{i + 2}}}}\underbrace {{{\cal L}^{ - 1}}\left[ {\frac{1}{{{{\left( {s + \lambda } \right)}^{k - i}}}}} \right]}_{ = \frac{{{t^{k - i - 1}}}}{{\left( {k - i - 1} \right)!}}{{\rm{e}}^{ - \lambda t}}}} 
 =  - \frac{{u\left( t \right)}}{{{\lambda ^{k + 1}}}} + \frac{{{{\rm{e}}^{\lambda t}}}}{{{2^k}{\lambda ^{k + 1}}}} + \sum\limits_{i = 0}^{k - 1} {\left( {1 - \frac{1}{{{2^{i + 1}}}}} \right){\lambda ^{ - \left( {i + 2} \right)}}\left[ {\frac{{{t^{k - i - 1}}}}{{\left( {k - i - 1} \right)!}}{{\rm{e}}^{ - \lambda t}}} \right]} 
\end{dmath}
\end{figure*}
Leveraging (\ref{eq54}), we now derive the inverse Laplace transforms of ${A_1},{A_2},{A_3},{A_4}$, resulting in (\ref{eq57})-(\ref{eq60}).
\begin{figure*}
\begin{align}
{{\cal L}^{ - 1}}\left[ {{A_1}} \right] &= {{\cal L}^{ - 1}}\left[ {\frac{{{\lambda ^{k + 1}}}}{{s\left( {s - \lambda } \right){{\left( {s + \lambda } \right)}^k}}}{{\rm{e}}^{ - s\left( {k - 1} \right){\tau _{\rm{d}}}}}} \right] \nonumber\\
 &= {\lambda ^{k + 1}}{{\cal L}^{ - 1}}\left[ {\frac{{{{\rm{e}}^{ - s\left( {k - 1} \right){\tau _{\rm{d}}}}}}}{{s\left( {s - \lambda } \right){{\left( {s + \lambda } \right)}^k}}}} \right] \label{eq57} \\ 
 &=  - u\left[ {t - (k - 1){\tau _{\rm{d}}}} \right] + {2^{ - k}}{{\rm{e}}^{\lambda \left[ {t - (k - 1){\tau _{\rm{d}}}} \right]}}
 + \sum\limits_{i = 0}^{k - 1} {\left( {1 - \frac{1}{{{2^{i + 1}}}}} \right){\lambda ^{k - \left( {i + 1} \right)}}\left[ {\frac{{{{\left[ {t - (k - 1){\tau _{\rm{d}}}} \right]}^{k - i - 1}}}}{{\left( {k - i - 1} \right)!}}{{\rm{e}}^{ - \lambda \left[ {t - (k - 1){\tau _{\rm{d}}}} \right]}}} \right]} \nonumber \\
{{\cal L}^{ - 1}}\left[ {{A_2}} \right] &= {{\cal L}^{ - 1}}\left[ {\frac{{{\lambda ^{k + 2}}}}{{s\left( {s - \lambda } \right){{\left( {s + \lambda } \right)}^{k + 1}}}}{{\rm{e}}^{ - sk{\tau _{\rm{d}}}}}} \right] \nonumber \\
 & = {\lambda ^{k + 2}}{{\cal L}^{ - 1}}\left[ {\frac{{{{\rm{e}}^{ - sk{\tau _{\rm{d}}}}}}}{{s\left( {s - \lambda } \right){{\left( {s + \lambda } \right)}^{k + 1}}}}} \right] \label{eq58} \\
&=  - u\left[ {t - k{\tau _{\rm{d}}}} \right] + {2^{ - \left( {k + 1} \right)}}{{\rm{e}}^{\lambda \left[ {t - k{\tau _{\rm{d}}}} \right]}}
 + \sum\limits_{i = 0}^k {\left( {1 - \frac{1}{{{2^{i + 1}}}}} \right){\lambda ^{k - i}}\left[ {\frac{{{{\left[ {t - k{\tau _{\rm{d}}}} \right]}^{k - i}}}}{{\left( {k - i} \right)!}}{{\rm{e}}^{ - \lambda \left[ {t - k{\tau _{\rm{d}}}} \right]}}} \right]}   \nonumber \\
 {{\cal L}^{ - 1}}\left[ {{A_3}} \right] &= {{\cal L}^{ - 1}}\left[ {\frac{{{\lambda ^{k + 1}}}}{{s\left( {s - \lambda } \right){{\left( {s + \lambda } \right)}^k}}}{{\rm{e}}^{ - sk{\tau _{\rm{d}}}}}} \right] \nonumber \\
 &= {\lambda ^{k + 1}}{{\cal L}^{ - 1}}\left[ {\frac{{{{\rm{e}}^{ - sk{\tau _{\rm{d}}}}}}}{{s\left( {s - \lambda } \right){{\left( {s + \lambda } \right)}^k}}}} \right] \label{eq59} \\
 &=  - u\left[ {t - k{\tau _{\rm{d}}}} \right] + {2^{ - k}}{{\rm{e}}^{\lambda \left[ {t - k{\tau _{\rm{d}}}} \right]}}
 + \sum\limits_{i = 0}^{k - 1} {\left( {1 - \frac{1}{{{2^{i + 1}}}}} \right){\lambda ^{k - \left( {i + 1} \right)}}\left[ {\frac{{{{\left[ {t - k{\tau _{\rm{d}}}} \right]}^{k - i - 1}}}}{{\left( {k - i - 1} \right)!}}{{\rm{e}}^{ - \lambda \left[ {t - k{\tau _{\rm{d}}}} \right]}}} \right]} \nonumber \\
 {{\cal L}^{ - 1}}\left[ {{A_4}} \right] &= {{\cal L}^{ - 1}}\left[ {\frac{{{\lambda ^{k + 2}}}}{{s\left( {s - \lambda } \right){{\left( {s + \lambda } \right)}^{k + 1}}}}{{\rm{e}}^{ - s\left( {k + 1} \right){\tau _{\rm{d}}}}}} \right] \nonumber \\
 &= {\lambda ^{k + 2}}{{\cal L}^{ - 1}}\left[ {\frac{{{{\rm{e}}^{ - s\left( {k + 1} \right){\tau _{\rm{d}}}}}}}{{s\left( {s - \lambda } \right){{\left( {s + \lambda } \right)}^{k + 1}}}}} \right] \label{eq60} \\
 &=  - u\left[ {t - \left( {k + 1} \right){\tau _{\rm{d}}}} \right] + {2^{ - \left( {k + 1} \right)}}{{\rm{e}}^{\lambda \left[ {t - \left( {k + 1} \right){\tau _{\rm{d}}}} \right]}}
 + \sum\limits_{i = 0}^k {\left( {1 - \frac{1}{{{2^{i + 1}}}}} \right){\lambda ^{k - i}}\left[ {\frac{{{{\left[ {t - \left( {k + 1} \right){\tau _{\rm{d}}}} \right]}^{k - i}}}}{{\left( {k - i} \right)!}}{{\rm{e}}^{ - \lambda \left[ {t - \left( {k + 1} \right){\tau _{\rm{d}}}} \right]}}} \right]} \nonumber   
\end{align}
\end{figure*}

Based on the inverse transforms given in (\ref{eq56})-(\ref{eq60}), the PMF of photon counts including ISI effects for the case $k < \left\lceil {{{{T_{\rm{s}}}} \mathord{\left/{\vphantom {{{T_{\rm{s}}}} {{\tau _{\rm{d}}}}}} \right. \kern-\nulldelimiterspace} {{\tau _{\rm{d}}}}}} \right\rceil$ is derived as (\ref{eq61}).
\begin{figure*}
\begin{dmath}\label{eq61}
p_{{\rm{ISI}}}^{\rm{L}}\left( {k,t} \right) = {{\cal L}^{ - 1}}\left[ {P_{{\rm{ISI}}}^{\rm{L}}\left( {k,s} \right)} \right] \nonumber\\
  = {{\cal L}^{ - 1}}\left[ B \right] + {{\rm{e}}^{ - \lambda {\tau _{\rm{d}}}}}\left( {{{\cal L}^{ - 1}}\left[ {{A_1}} \right] - {{\cal L}^{ - 1}}\left[ {{A_2}} \right]} \right) - \left( {{{\cal L}^{ - 1}}\left[ {{A_3}} \right] - {{\cal L}^{ - 1}}\left[ {{A_4}} \right]} \right)\nonumber\\
 = \left\langle {\sum\limits_{i = 0}^k {\frac{{{\lambda ^i}}}{{i!}}{{\left[ {t - k{\tau _{\rm{d}}}} \right]}^i}{{\rm{e}}^{ - \lambda \left[ {t - \left( {k - 1} \right){\tau _{\rm{d}}}} \right]}}}  - \sum\limits_{i = 0}^{k - 1} {\frac{{{\lambda ^i}}}{{i!}}{{\left[ {t - \left( {k - 1} \right){\tau _{\rm{d}}}} \right]}^i}{{\rm{e}}^{ - \lambda \left[ {t - \left( {k - 2} \right){\tau _{\rm{d}}}} \right]}}} } \right\rangle \nonumber\\
  + \left\langle \sum\limits_{i = 0}^{k - 1} {\left( {1 - \frac{1}{{{2^{k - i}}}}} \right)\frac{{{\lambda ^i}}}{{i!}}{{\left[ {t - \left( {k - 1} \right){\tau _{\rm{d}}}} \right]}^i}{{\rm{e}}^{ - \lambda \left[ {t - \left( {k - 2} \right){\tau _{\rm{d}}}} \right]}} - \sum\limits_{i = 0}^{k - 1} {\left( {1 - \frac{1}{{{2^{k - i}}}}} \right)\frac{{{\lambda ^i}}}{{i!}}{{\left[ {t - k{\tau _{\rm{d}}}} \right]}^i}{{\rm{e}}^{ - \lambda \left[ {t - k{\tau _{\rm{d}}}} \right]}}} } \right\rangle \nonumber\\
  + \left\langle \sum\limits_{i = 0}^k {\left( {1 - \frac{1}{{{2^{k - i + 1}}}}} \right)\frac{{{\lambda ^i}}}{{i!}}{{\left[ {t - \left( {k + 1} \right){\tau _{\rm{d}}}} \right]}^i}{{\rm{e}}^{ - \lambda \left[ {t - \left( {k + 1} \right){\tau _{\rm{d}}}} \right]}}}  - \sum\limits_{i = 0}^k {\left( {1 - \frac{1}{{{2^{k - i + 1}}}}} \right)\frac{{{\lambda ^i}}}{{i!}}{{\left[ {t - k{\tau _{\rm{d}}}} \right]}^i}{{\rm{e}}^{ - \lambda \left[ {t - \left( {k - 1} \right){\tau _{\rm{d}}}} \right]}}}  \right\rangle \nonumber
\end{dmath}    
\end{figure*}

The expression for the PMF becomes computationally intractable when $k$ reaches its theoretical maximum value. To address this, we introduce an approximation valid only when ISI is negligible. Under this ISI-free assumption ($k = \left\lceil {{{{T_{\rm{s}}}} \mathord{\left/
 {\vphantom {{{T_{\rm{s}}}} {{\tau _{\rm{d}}}}}} \right.
 \kern-\nulldelimiterspace} {{\tau _{\rm{d}}}}}} \right\rceil$), the PMF for $k = {k_{\max }}$ is approximated as:
\begin{align}\label{eq62}
p_{{\rm{ISI}}}^{\rm{L}}\left( {k,t} \right) &\ge {{\rm{e}}^{ - \lambda {\tau _{\rm{d}}}}}p_{{\rm{NISI}}}^{\rm{L}}\left( {\left\lceil {{{{T_{\rm{s}}}} \mathord{\left/
 {\vphantom {{{T_{\rm{s}}}} {{\tau _{\rm{d}}}}}} \right.
 \kern-\nulldelimiterspace} {{\tau _{\rm{d}}}}}} \right\rceil ,t} \right) \\
&= {{\rm{e}}^{ - \lambda {\tau _{\rm{d}}}}}\left\langle {1 - \sum\limits_{i = 0}^{k - 1} {\frac{{{\lambda ^i}}}{{i!}}{{\left[ {t - \left( {k - 1} \right){\tau _{\rm{d}}}} \right]}^i}{{\rm{e}}^{ - \lambda \left[ {t - \left( {k - 1} \right){\tau _{\rm{d}}}} \right]}}} } \right\rangle \nonumber\\
&= {{\rm{e}}^{ - \lambda {\tau _{\rm{d}}}}} - \sum\limits_{i = 0}^{k - 1} {\frac{{{\lambda ^i}}}{{i!}}{{\left[ {t - \left( {k - 1} \right){\tau _{\rm{d}}}} \right]}^i}{{\rm{e}}^{ - \lambda \left[ {t - \left( {k - 2} \right){\tau _{\rm{d}}}} \right]}}}  \nonumber  
\end{align}

The equality only holds under the ${{{T_{\rm{s}}}} \mathord{\left/ {\vphantom {{{T_{\rm{s}}}} {{\tau _{\rm{d}}}}}} \right. \kern-\nulldelimiterspace} {{\tau _{\rm{d}}}}}$ is an integer. However, within the operating regimes considered in this paper, ${{{T_{\rm{s}}}} \mathord{\left/{\vphantom {{{T_{\rm{s}}}} {{\tau _{\rm{d}}}}}} \right. \kern-\nulldelimiterspace} {{\tau _{\rm{d}}}}}$ predominantly takes integer values. Therefore, employing this approximation is both feasible and justified for our analysis.

Combining the exact PMF for $k < {k_{\max }}$ (\ref{eq61}) with the approximation for $k = {k_{\max }}$ (\ref{eq62}) provides the complete description of the photon counts incorporating ISI effects under the specified conditions. This completes the derivation.

\section*{Appendix D \\PDF of ${T_{{\rm{ISI}}}}$ for High-Speed Systems ($\xi \ge 1$)}
Firstly, substituting (\ref{eq18}) into (\ref{eq19}), yields the PDF of ${T_{{\rm{ISI}}}}$ for the second symbol (\ref{eq63}). Next, substituting (\ref{eq63}) into (\ref{eq19}), yields the PDF of ${T_{{\rm{ISI}}}}$ for the third symbol (\ref{eq64}). Following the same methodology, the PDFs of ${T_{{\rm{ISI}}}}$ for the fourth, fifth, and sixth symbols are derived as (\ref{eq65})-(\ref{eq67}).
\begin{figure*}
\begin{align}
f_{{T_{{\rm{ISI}}}}}^2\left( t \right) &= \int_0^{{T_{\rm{s}}}} {\left\{ {\lambda {{\rm{e}}^{ - \lambda \left( {t - t'} \right)}}\left[ {u\left( {t - t'} \right) - u\left( {t - t' - {T_{\rm{s}}}} \right)} \right] + {{\rm{e}}^{ - \lambda \left( {{T_{\rm{s}}} - t'} \right)}}\delta \left( t \right)} \right\}f_{{T_{{\rm{ISI}}}}}^1\left( {t'} \right)dt'} \label{eq63} \\
&= \int_0^{{T_{\rm{s}}}} {\left\{ {\lambda {{\rm{e}}^{ - \lambda \left( {t - t'} \right)}}\left[ {u\left( {t - t'} \right) - u\left( {t - t' - {T_{\rm{s}}}} \right)} \right] + {{\rm{e}}^{ - \lambda \left( {{T_{\rm{s}}} - t'} \right)}}\delta \left( t \right)} \right\}\left\{ {\lambda {{\rm{e}}^{ - \lambda t'}}\left[ {u\left( {t'} \right) - u\left( {t' - {T_{\rm{s}}}} \right)} \right] + {{\rm{e}}^{ - \lambda {T_{\rm{s}}}}}\xi \left( {t'} \right)} \right\}dt'} \nonumber\\
&= \int_0^t {{\lambda ^2}{{\rm{e}}^{ - \lambda t}}} dt' + \int_0^{{T_{\rm{s}}}} {\lambda {{\rm{e}}^{ - \lambda {T_{\rm{s}}}}}} dt'\delta \left( t \right) + {\left. {{{\rm{e}}^{ - \lambda {T_{\rm{s}}}}}\left[ {\lambda {{\rm{e}}^{ - \lambda \left( {t - t'} \right)}} + {{\rm{e}}^{ - \lambda \left( {{T_{\rm{s}}} - t'} \right)}}\delta \left( t \right)} \right]} \right|_{t' = 0}} \nonumber\\
&= \lambda {{\rm{e}}^{ - \lambda t}}\left[ {\lambda t + {{\rm{e}}^{ - \lambda {T_{\rm{s}}}}}} \right]\left[ {u\left( t \right) - u\left( {t - {T_{\rm{s}}}} \right)} \right] + {{\rm{e}}^{ - \lambda {T_{\rm{s}}}}}\left[ {\lambda {T_{\rm{s}}} + {{\rm{e}}^{ - \lambda {T_{\rm{s}}}}}} \right]\delta \left( t \right)  \nonumber \\
f_{{T_{{\rm{ISI}}}}}^3\left( t \right) &= \int_0^{{T_{\rm{s}}}} {\left\{ {\lambda {{\rm{e}}^{ - \lambda \left( {t - t'} \right)}}\left[ {u\left( {t - t'} \right) - u\left( {t - t' - {T_{\rm{s}}}} \right)} \right] + {{\rm{e}}^{ - \lambda \left( {{T_{\rm{s}}} - t'} \right)}}\delta \left( t \right)} \right\}f_{{T_{{\rm{ISI}}}}}^2\left( {t'} \right)dt'} \label{eq64} \\
&= \lambda {{\rm{e}}^{ - \lambda t}}\left[ {\frac{{{{\left( {\lambda t} \right)}^2}}}{2} + \left( {\lambda {T_{\rm{s}}} + \lambda t} \right){{\rm{e}}^{ - \lambda {T_{\rm{s}}}}}{\rm{ + }}{{\rm{e}}^{ - 2\lambda {T_{\rm{s}}}}}} \right]\left[ {u\left( t \right) - u\left( {t - {T_{\rm{s}}}} \right)} \right] \nonumber \\
&\hspace{1em} + {{\rm{e}}^{ - \lambda {T_{\rm{s}}}}}\left[ {\frac{{{{\left( {\lambda {T_{\rm{s}}}} \right)}^2}}}{2} + 2\lambda {T_{\rm{s}}}{{\rm{e}}^{ - \lambda {T_{\rm{s}}}}}{\rm{ + }}{{\rm{e}}^{ - 2\lambda {T_{\rm{s}}}}}} \right]\delta \left( t \right) \nonumber \\
f_{{T_{{\rm{ISI}}}}}^4\left( t \right) &= \int_0^{{T_{\rm{s}}}} {\left\{ {\lambda {{\rm{e}}^{ - \lambda \left( {t - t'} \right)}}\left[ {u\left( {t - t'} \right) - u\left( {t - t' - {T_{\rm{s}}}} \right)} \right] + {{\rm{e}}^{ - \lambda \left( {{T_{\rm{s}}} - t'} \right)}}\delta \left( t \right)} \right\}f_{{T_{{\rm{ISI}}}}}^3\left( {t'} \right)dt'} \label{eq65} \\
&= \lambda {{\rm{e}}^{ - \lambda t}}\left[ {\frac{{{{\left( {\lambda t} \right)}^3}}}{{3}} + \frac{{{{\left( {\lambda t} \right)}^2} + 2{\lambda ^2}t{T_{\rm{s}}} + {{\left( {\lambda {T_{\rm{s}}}} \right)}^2}}}{2}{{\rm{e}}^{ - \lambda {T_{\rm{s}}}}}{\rm{ + }}\left( {\lambda t + 2\lambda {T_{\rm{s}}}} \right){{\rm{e}}^{ - 2\lambda {T_{\rm{s}}}}} + {{\rm{e}}^{ - 3\lambda {T_{\rm{s}}}}}} \right]\left[ {u\left( t \right) - u\left( {t - {T_{\rm{s}}}} \right)} \right] \nonumber \\
&\hspace{1em} + {{\rm{e}}^{ - \lambda {T_{\rm{s}}}}}\left[ {\frac{{{{\left( {\lambda {T_{\rm{s}}}} \right)}^3}}}{{3}} + 2{{\left( {\lambda {T_{\rm{s}}}} \right)}^2}{{\rm{e}}^{ - \lambda {T_{\rm{s}}}}}{\rm{ + }}3\lambda {T_{\rm{s}}}{{\rm{e}}^{ - 2\lambda {T_{\rm{s}}}}} + {{\rm{e}}^{ - 3\lambda {T_{\rm{s}}}}}} \right]\delta \left( t \right) \nonumber \\
f_{{T_{{\rm{ISI}}}}}^5\left( t \right) &= \int_0^{{T_{\rm{s}}}} {\left\{ {\lambda {{\rm{e}}^{ - \lambda \left( {t - t'} \right)}}\left[ {u\left( {t - t'} \right) - u\left( {t - t' - {T_{\rm{s}}}} \right)} \right] + {{\rm{e}}^{ - \lambda \left( {{T_{\rm{s}}} - t'} \right)}}\delta \left( t \right)} \right\}f_{{T_{{\rm{ISI}}}}}^4\left( {t'} \right)dt'} \label{eq66} \\
&= \lambda {{\rm{e}}^{ - \lambda t}}\left[ { \begin{array}{l} \frac{{2{{\left( {\lambda t} \right)}^3} + 3{\lambda ^3}{t^2}{T_{\rm{s}}} + {{\left( {\lambda {T_{\rm{s}}}} \right)}^3}}}{6}{{\rm{e}}^{ - \lambda {T_{\rm{s}}}}} + \frac{{{{\left( {\lambda t} \right)}^2} + 4{\lambda ^2}t{T_{\rm{s}}}}}{2}{{\rm{e}}^{ - 2\lambda {T_{\rm{s}}}}} \\ 
+ \left( {\lambda t + 3\lambda {T_{\rm{s}}}} \right){{\rm{e}}^{ - 3\lambda {T_{\rm{s}}}}}{\rm{ + }}{{\rm{e}}^{ - 4\lambda {T_{\rm{s}}}}} +  \frac{{{{\left( {\lambda t} \right)}^4}}}{{4}} \end{array}} \right]\left[ {u\left( t \right) - u\left( {t - {T_{\rm{s}}}} \right)} \right]\nonumber\\
&\hspace{1em} + {{\rm{e}}^{ - \lambda {T_{\rm{s}}}}}\left[ {\frac{{{{\left( {\lambda {T_{\rm{s}}}} \right)}^4}}}{{4}} + {{\left( {\lambda {T_{\rm{s}}}} \right)}^3}{{\rm{e}}^{ - \lambda {T_{\rm{s}}}}}{\rm{ + }}\frac{{9{{\left( {\lambda {T_{\rm{s}}}} \right)}^2}}}{2}{{\rm{e}}^{ - 2\lambda {T_{\rm{s}}}}} + 4\lambda {T_{\rm{s}}}{{\rm{e}}^{ - 3\lambda {T_{\rm{s}}}}}{\rm{ + }}{{\rm{e}}^{ - 4\lambda {T_{\rm{s}}}}}} \right]\delta \left( t \right)  \nonumber \\
f_{{T_{{\rm{ISI}}}}}^6\left( t \right)&= \int_0^{{T_{\rm{s}}}} {\left\{ {\lambda {{\rm{e}}^{ - \lambda \left( {t - t'} \right)}}\left[ {u\left( {t - t'} \right) - u\left( {t - t' - {T_{\rm{s}}}} \right)} \right] + {{\rm{e}}^{ - \lambda \left( {{T_{\rm{s}}} - t'} \right)}}\delta \left( t \right)} \right\}f_{{T_{{\rm{ISI}}}}}^5\left( {t'} \right)dt'} \label{eq67} \\
&= \lambda {{\rm{e}}^{ - \lambda t}}\left\{ \begin{array}{l}
 \frac{{2{{\left( {\lambda t} \right)}^4} + 4\left( {{\lambda ^4}{t^3}{T_{\rm{s}}} + {\lambda ^4}t{T_{\rm{s}}}^3} \right) + {{\left( {\lambda {T_{\rm{s}}}} \right)}^4}}}{{24}}{{\rm{e}}^{ - \lambda {T_{\rm{s}}}}}\\
 + \frac{{{{\left( {\lambda t} \right)}^3} + 6{\lambda ^3}{t^2}{T_{\rm{s}}} + 12{\lambda ^3}t{T_{\rm{s}}}^2 + 6{{\left( {\lambda {T_{\rm{s}}}} \right)}^3}}}{6}{{\rm{e}}^{ - 2\lambda {T_{\rm{s}}}}}\\
 + \frac{{{{\left( {\lambda t} \right)}^2} + 6{\lambda ^2}t{T_{\rm{s}}} + 9{{\left( {\lambda {T_{\rm{s}}}} \right)}^2}}}{2}{{\rm{e}}^{ - 3\lambda {T_{\rm{s}}}}}\\
 + \frac{{\lambda t + 4\lambda {T_{\rm{s}}}}}{2}{{\rm{e}}^{ - 4\lambda {T_{\rm{s}}}}} + {{\rm{e}}^{ - 5\lambda {T_{\rm{s}}}}}
 + \frac{{{{\left( {\lambda t} \right)}^5}}}{{120}} 
\end{array} \right\}  \left[  {u\left( t \right)  - u\left( {t - {T_{\rm{s}}}} \right)} \right]
+ \left\{ \begin{array}{l}
\frac{{{{\left( {\lambda {T_{\rm{s}}}} \right)}^5}}}{{120}}{{\rm{e}}^{ - \lambda {T_{\rm{s}}}}} \\ + \frac{{11{{\left( {\lambda {T_{\rm{s}}}} \right)}^4}}}{{24}}{{\rm{e}}^{ - 2\lambda {T_{\rm{s}}}}}\\
 + \frac{{25{{\left( {\lambda {T_{\rm{s}}}} \right)}^3}}}{6}{{\rm{e}}^{ - 3\lambda {T_{\rm{s}}}}}\\
 + 8{\left( {\lambda {T_{\rm{s}}}} \right)^2}{{\rm{e}}^{ - 4\lambda {T_{\rm{s}}}}}\\
 + 5\lambda {T_{\rm{s}}}{{\rm{e}}^{ - 5\lambda {T_{\rm{s}}}}}  \\ +{{\rm{e}}^{ - 6\lambda {T_{\rm{s}}}}}
\end{array} \right\}\delta \left( t \right)  \nonumber
\end{align}   
\end{figure*}

This completes the derivation of the PDFs.
\FloatBarrier

\section*{Appendix E \\Steady-State Probability Distribution}
In order to solve the linear equations (\ref{eq25}), a linear transformation is applied to the coefficient matrix as shown in (\ref{eq68}).
\begin{figure*}
\begin{dmath}\label{eq68}
\hspace{2em} \left[ {\begin{array}{*{20}{c}}
{{{\tilde p}_1} - 1}&0&0& \cdots &0&{{p_1}}\\
{{{\tilde p}_0}{p_1}}&{{{\tilde p}_1} - 1}&0& \cdots &0&{{p_1}{p_0}}\\
{{{\tilde p}_0}{p_1}{p_0}}&{{{\tilde p}_0}{p_1}}&{{{\tilde p}_1} - 1}& \cdots &0&{{p_1}{p_0}{}^2}\\
 \vdots & \vdots & \vdots & \ddots & \vdots & \vdots \\
{{{\tilde p}_0}{p_1}{p_0}{}^{\xi  - 2}}&{{{\tilde p}_0}{p_1}{p_0}{}^{\xi  - 3}}&{{{\tilde p}_0}{p_1}{p_0}{}^{\xi  - 4}}& \cdots &{{{\tilde p}_1} - 1}&{{p_1}{p_0}{}^{\xi  - 1}}\\
{{{\tilde p}_0}{p_0}{}^{\xi  - 1}}&{{{\tilde p}_0}{p_0}{}^{\xi  - 2}}&{{{\tilde p}_0}{p_0}{}^{\xi  - 3}}& \cdots &{{{\tilde p}_0}}&{{p_0}{}^\xi  - 1}\\
1&1&1& \cdots &1&1
\end{array}} \right] \\
\to \left[ {\begin{array}{*{20}{c}}
{ - {{\tilde p}_0}}&0&0& \cdots &0&{{p_1}}\\
{{{\tilde p}_0}}&{ - {{\tilde p}_0}}&0& \cdots &0&0\\
{{{\tilde p}_0}{p_1}{p_0}}&{{{\tilde p}_0}{p_1}}&{ - {{\tilde p}_0}}& \cdots &0&{{p_1}{p_0}{}^2}\\
 \vdots & \vdots & \vdots & \ddots & \vdots & \vdots \\
{{{\tilde p}_0}{p_1}{p_0}{}^{\xi  - 2}}&{{{\tilde p}_0}{p_1}{p_0}{}^{\xi  - 3}}&{{{\tilde p}_0}{p_1}{p_0}{}^{\xi  - 4}}& \cdots &{ - {{\tilde p}_0}}&{{p_1}{p_0}{}^{\xi  - 1}}\\
{{{\tilde p}_0}{p_0}{}^{\xi  - 1}}&{{{\tilde p}_0}{p_0}{}^{\xi  - 2}}&{{{\tilde p}_0}{p_0}{}^{\xi  - 3}}& \cdots &{{{\tilde p}_0}}&{{p_0}{}^\xi  - 1}\\
1&1&1& \cdots &1&1
\end{array}} \right] \\
\to \left[ {\begin{array}{*{20}{c}}
{ - {{\tilde p}_0}}&0&0& \cdots &0&{{p_1}}\\
{{{\tilde p}_0}}&{ - {{\tilde p}_0}}&0& \cdots &0&0\\
{{{\tilde p}_0}}&0&{ - {{\tilde p}_0}}& \cdots &0&0\\
 \vdots & \vdots & \vdots & \ddots & \vdots & \vdots \\
{{{\tilde p}_0}{p_1}{p_0}{}^{\xi  - 2}}&{{{\tilde p}_0}{p_1}{p_0}{}^{\xi  - 3}}&{{{\tilde p}_0}{p_1}{p_0}{}^{\xi  - 4}}& \cdots &{ - {{\tilde p}_0}}&{{p_1}{p_0}{}^{\xi  - 1}}\\
{{{\tilde p}_0}{p_0}{}^{\xi  - 1}}&{{{\tilde p}_0}{p_0}{}^{\xi  - 2}}&{{{\tilde p}_0}{p_0}{}^{\xi  - 3}}& \cdots &{{{\tilde p}_0}}&{{p_0}{}^\xi  - 1}\\
1&1&1& \cdots &1&1
\end{array}} \right] \to \cdots \cdots \\
 \to \left[ {\begin{array}{*{20}{c}}
{ - {{\tilde p}_0}}&0&0& \cdots &0&{{p_1}}\\
{{{\tilde p}_0}}&{ - {{\tilde p}_0}}&0& \cdots &0&0\\
{{{\tilde p}_0}}&0&{ - {{\tilde p}_0}}& \cdots &0&0\\
 \vdots & \vdots & \vdots & \ddots & \vdots & \vdots \\
{{{\tilde p}_0}}&0&0& \cdots &{ - {{\tilde p}_0}}&0\\
{{{\tilde p}_0}{p_0}{}^{\xi  - 1}}&{{{\tilde p}_0}{p_0}{}^{\xi  - 2}}&{{{\tilde p}_0}{p_0}{}^{\xi  - 3}}& \cdots &{{{\tilde p}_0}}&{{p_0}{}^\xi  - 1}\\
1&1&1& \cdots &1&1
\end{array}} \right]    
\end{dmath}   
\end{figure*}
Based on (\ref{eq68}), it follows that $\gamma \left( 1 \right) = \gamma \left( 2 \right) =  \cdots  = \gamma \left( \xi  \right)$. Then, utilizing $\sum\nolimits_{n = 0}^{\xi  + 1} {\gamma \left( n \right)}  = 1$, we obtain $\gamma \left( {\xi  + 1} \right) = 1 - \xi \gamma \left( 1 \right)$. Substituting these two expressions into the first linear equation,
\begin{align}\label{eq69}
 - {\tilde p_0}\gamma \left( 1 \right) + {p_1}\left( {1 - \xi \gamma \left( 1 \right)} \right) = \gamma \left( 1 \right) 
\end{align}

Then, it yields the unique solution to the system of linear equations, given as follows:
\begin{align}\label{eq70}
\gamma \left( 1 \right) = \frac{{{p_1}}}{{\xi {p_1} + {{\tilde p}_0}}} 
\end{align}

Consequently, the last steady-state probability is 
\begin{align}\label{eq71}
\gamma \left( {\xi  + 1} \right) = \frac{{{{\tilde p}_0}}}{{\xi {p_1} + {{\tilde p}_0}}}  
\end{align}

This completes the proof.

\FloatBarrier
 
%

\vfill

\end{document}